\documentclass[a4paper,12pt]{article}
\pdfoutput=1
\usepackage{graphicx, rotating}
\usepackage{hyperref}
\usepackage{slashed}
\usepackage{ifpdf}

\ifx\pdfoutput\undefined
\usepackage[dvips,bookmarks=false]{hyperref}	
\else
\usepackage{hyperref}	
\fi
\hypersetup{colorlinks,bookmarksopen,bookmarksnumbered,citecolor=verdes,
linkcolor=blus,pdfstartview=FitH,urlcolor=rossos}
\def\hhref#1{\href{http://arxiv.org/abs/#1}{#1}} 

\usepackage{amsmath}
\usepackage{slashed}

\allowdisplaybreaks

\newcommand{\beq}{\begin{equation}}
\newcommand{\eeq}{\end{equation}}
\newcommand{\fig}[1]{~\ref{fig:#1}}

\newcount\Mac  \Mac=1  
\newcommand{\ifMac}[2]{\ifnum\Mac=1 #1 \else #2 \fi}
\def\putps(#1,#2)(#3,#4)#5#6{\ifnum\Mac=1 \put(#1,#2){\special{picture #5}}
\else  \put(#3,#4){\includegraphics{#6}} \fi}

\newcommand{\One}{\hbox{1\kern-.24em I}}

\newcommand{\GeV}{\,{\rm GeV}}

\newcommand{\eq}[1]{~{\rm(\ref{eq:#1})}}

\newcommand{\lascia}[1]{}
\makeatletter
\def\art{\@ifnextchar[{\eart}{\oart}}
\def\eart[#1]#2#3#4#5#6{{\rm #2}, {#3 #4} {\rm (#6) #5} [arXiv:{\hhref{#1}}]}
\def\hepart[#1]#2{{\rm #2, arXiv:\hhref{#1}}}
\newcommand{\oart}[5]{{\rm #1}, {#2 #3} {\rm (#5) #4}}

\newcounter{alphaequation}[equation]
\def\thealphaequation{\theequation\hbox to
0.6em{\hfil\alph{alphaequation}\hfil}}
\def\eqnsystem#1{
\def\@eqnnum{{\rm (\thealphaequation)}}
\def\@@eqncr{\let\@tempa\relax \ifcase\@eqcnt \def\@tempa{& & &} \or
  \def\@tempa{& &}\or \def\@tempa{&}\fi\@tempa
  \if@eqnsw\@eqnnum\refstepcounter{alphaequation}\fi
\global\@eqnswtrue\global\@eqcnt=0\cr}
\refstepcounter{equation} \let\@currentlabel\theequation \def\@tempb{#1}
\ifx\@tempb\empty\else\label{#1}\fi
\refstepcounter{alphaequation}
\let\@currentlabel\thealphaequation
\global\@eqnswtrue\global\@eqcnt=0 \tabskip\@centering\let\\=\@eqncr
$$\halign to \displaywidth\bgroup \@eqnsel\hskip\@centering
$\displaystyle\tabskip\z@{##}$&\global\@eqcnt\@ne
\hskip2\arraycolsep\hfil${##}$\hfil& \global\@eqcnt\tw@\hskip2\arraycolsep
$\displaystyle\tabskip\z@{##}$\hfil
\tabskip\@centering&\llap{##}\tabskip\z@\cr}

\def\endeqnsystem{\@@eqncr\egroup$$\global\@ignoretrue} \makeatother

\def\Lag{{\cal L}}
\def\SU{{\rm SU}}

\def\Tr{\mathop{\rm Tr}}
\def\circa#1{\,\raise.3ex\hbox{$#1$\kern-.75em\lower1ex\hbox{$\sim$}}\,}

\usepackage{multicol}
\usepackage{color}
\definecolor{rosso}{cmyk}{0,1,1,0.4}
\definecolor{rossos}{cmyk}{0,1,1,0.55}
\definecolor{rossoc}{cmyk}{0,1,1,0.2}
\definecolor{blu}{cmyk}{1,1,0,0.3}
\definecolor{blus}{cmyk}{1,1,0,0.6}
\definecolor{bluc}{cmyk}{1,1,0,0.1}
\definecolor{verde}{cmyk}{0.92,0,0.59,0.25}
\definecolor{verdec}{cmyk}{0.92,0,0.59,0.15}
\definecolor{verdes}{cmyk}{0.92,0,0.59,0.4}
\definecolor{grigio}{cmyk}{0,0,0,0.07}
\definecolor{rosa}{cmyk}{0,0.1,0.1,0.02}
\definecolor{rosino}{cmyk}{0,0.05,0.05,0.02}
\definecolor{rosas}{cmyk}{0,0.3,0.25,0.05}
\definecolor{celeste}{cmyk}{0.1,0,0,0.02}
\definecolor{giallino}{cmyk}{0,0,0.4,0.02}
\definecolor{rosso}{cmyk}{0,1,1,0.4}
\definecolor{rossos}{cmyk}{0,1,1,0.55}
\definecolor{rossoc}{cmyk}{0,1,1,0.2}
\definecolor{blu}{cmyk}{1,1,0,0.3}
\definecolor{bluc}{cmyk}{1,1,0,0.1}
\definecolor{blucc}{cmyk}{0.7,0.5,0,0}
\definecolor{viola}{cmyk}{0,1,0,0.6}
\definecolor{viola2}{cmyk}{0,1,0.2,0.6}
\definecolor{verde}{cmyk}{0.92,0,0.59,0.25}
\definecolor{verdec}{cmyk}{0.92,0,0.59,0.15}
\definecolor{verdes}{cmyk}{0.92,0,0.59,0.4}
\definecolor{verdino}{cmyk}{0.12,0,0.09,0.05}
\definecolor{giallo}{cmyk}{0,0,1,0}
\definecolor{gialloverde}{cmyk}{0.44,0,0.74,0}

\font\tenrsfs=rsfs10 at 12pt

\font\sevenrsfs=rsfs7
\font\fiversfs=rsfs5
\newfam\rsfsfam
\textfont\rsfsfam=\tenrsfs
\scriptfont\rsfsfam=\sevenrsfs
\scriptscriptfont\rsfsfam=\fiversfs
\def\mathscr#1{{\fam\rsfsfam\relax#1}}
\def\Lag{\mathscr{L}}

\def\order#1{{\cal O}(#1)}
\def\eq#1{eq.~(\ref{#1})}
\def\beq{\begin{equation}}
\def\eeq{\end{equation}}
\def\bea{\begin{eqnarray}}
\def\eea{\end{eqnarray}}
\def\tm{{\tilde m}}

\begin{document}\hfill
 IFUP-TH/2011\hfill 
 CERN-PH-TH/2011-215\hfill 
\color{black}
\vspace{1cm}
\begin{center}
{\LARGE\bf\color{black}Probing High-Scale and Split Supersymmetry\\[3mm] with Higgs Mass Measurements}\\
\bigskip\color{black}\vspace{0.6cm}{
{\large\bf Gian F.\ Giudice$^{a}$ {\rm and} Alessandro Strumia$^{b,c}$}
} \\[7mm]
{\it (a) CERN, Theory Division, CH-1211 Geneva 23, Switzerland}\\[1mm]
{\it (b) Dipartimento di Fisica dell'Universit{\`a} di Pisa and INFN, Italy}\\[1mm]
{\it  (c) National Institute of Chemical Physics and Biophysics, Ravala 10, Tallinn, Estonia}\\[3mm]
\end{center}
\bigskip
\centerline{\large\bf\color{blus} Abstract}

\begin{quote}\large
We study the range of Higgs masses predicted by High-Scale Supersymmetry and by Split Supersymmetry, using the matching condition for the Higgs quartic coupling determined by the minimal field content. In the case of Split Supersymmetry, we compute for the first time the complete next-to-leading order corrections, including two-loop
renormalization group equations and one loop threshold effects. These corrections reduce
the predicted Higgs mass by a few GeV. We investigate the impact of the recent LHC Higgs searches on the scale of supersymmetry breaking. In particular, we show that an upper bound of 127 GeV on the Higgs mass implies an upper bound on the scale of Split Supersymmetry of about $10^8\GeV$, while no firm conclusion can yet be drawn for High-Scale Supersymmetry. \end{quote}


\section{Introduction}

Supersymmetry is often considered one of the most appealing theoretical ingredients for extending the Standard Model of particle physics. The interest on supersymmetry is based on four main reasons: {\it (i)} solution of the naturalness problem, {\it (ii)} successful gauge coupling unification, {\it (iii)} viable thermal dark matter candidate, {\it (iv)} necessary element of string theory. 

Only the first three reasons establish
a link between supersymmetry and the weak scale while, as far as {\it (iv)} is concerned, supersymmetry could be broken at any scale below the Planck mass. 

However, it is fair to say that only {\it (i)} provides a firm reason to believe  that supersymmetry should be discovered at the LHC. Points {\it (ii)} and {\it (iii)}, although linking supersymmetry to the weak scale, do not necessarily guarantee discoveries at the LHC. Indeed, gauge coupling unification, being only logarithmically sensitive to the supersymmetric mass scale, is consistent (and actually even more successful) with superparticles in the multi-TeV range. Thermal dark matter can be obtained with higgsinos as heavy as 1 TeV or winos of 2.7 TeV~\cite{Hisano}, without setting any constraint on new colored particles. As a result, both gauge coupling unification and dark matter can be successfully obtained in many simple supersymmetric schemes that do not predict any new particles within reach of the LHC.

Naturalness has been for long a guiding principle for constructing theories beyond the Standard Model. Given the remarkable successes of our understanding of the particle world in terms of effective theories, naturalness looks like a very plausible lead in the search for new physics. Nonetheless, the naturalness criterion represents only a valid theoretical argument, but not a logical necessity. The rising degree of fine tuning present in supersymmetric models~\cite{ftlhc}, the difficulty in dealing with the cosmological constant, and the theoretical exploration of the multiverse have all contributed in creating a certain amount of skepticism among physicists about naturalness arguments. 

A drastic departure from the conventional paradigm is given by the interesting possibility of Split Supersymmetry~\cite{split,GR,GR4}, in which one retains the motivations in {\it (ii)}--{\it (iv)}, but abandons {\it (i)}. As a result of an approximate $R$-symmetry or of an accidental symmetry arising from the pattern of supersymmetry breaking (as in the case of $D$-term breaking), dimension-2 operators corresponding to scalar masses are generated at a high scale, while dimension-3 operators for higgsino and gaugino masses, and for trilinear $A$-terms have only weak-scale size. Lacking the strong constraint from naturalness, there is no guarantee that Split Supersymmetry will be discovered at the LHC, even if it is indeed realized in nature. The chance of discovery are tied to the existence of gluinos lighter than 2.5--3.0 TeV or of charginos and neutralinos below several hundreds of GeV. Unification and dark matter do not necessarily favor this situation. 

An even more extreme option is using only {\it (iv)} as guidance, surrendering any link between supersymmetry and the weak scale. We have in mind a situation in which all supersymmetric partners have roughly equal masses at a high scale $\tm$. We will refer to this case as High-Scale Supersymmetry. Naively it may seem that High-Scale Supersymmetry, although interesting in the context of quantum gravity and string theory, has absolutely no chance of being probed at collider experiments. This is not necessarily the case.

Measurements of the Higgs mass can provide quite useful information on a putative high scale of supersymmetry breaking or set strong constraints on its existence. Let us consider the case in which, below the scale $\tm$, the effective theory contains only the Standard Model degrees of freedom (for High-Scale Supersymmetry) or additional fermions (for Split Supersymmetry) needed for gauge coupling unification and dark matter. The information about the matching with a full supersymmetric theory is encoded in the Higgs quartic coupling. Thus a measurement of the Higgs mass can probe the existence of supersymmetry, even if the symmetry is broken at a high scale.

In this paper we perform a detailed analysis of the Higgs mass predictions in models with a high scale of supersymmetry breaking, discussing the sources of theoretical uncertainties. Previous results for the case of Split Supersymmetry were presented in ref.~\cite{split,GR,GR4,Binger} and, more recently, in ref.~\cite{new}. The case of High-Scale Supersymmetry was studied in ref.~\cite{nomura}. 
\begin{itemize}
\item In section~\ref{RGEs} we present the full next-to-leading order analysis, including the two-loop calculation of the Renormalization Group Equations (RGE) for the couplings of Split Supersymmetry and the one-loop threshold effects. The two-loop calculation of the RGE for the Higgs quartic coupling has already been presented in ref.~\cite{Binger}. 
We extend the calculation to the RGE of the top Yukawa and the gaugino-higgsino-Higgs couplings,
which are needed for a consistent two-loop prediction of the Higgs mass.
 We find that the most sizable two-loop contribution to the Higgs mass comes from the newly calculated top Yukawa RGE, rather than from the Higgs quartic coupling. For High-Scale Supersymmetry we study how the Higgs mass prediction changes as the supersymmetry breaking mass is varied, including threshold effects. 
 \item In section~\ref{hi} we present the one loop matching conditions at the high scale.
 \item In section~\ref{low} we summarize the one loop matching conditions at the high scale due to SM particles and,
 in the case of Split Supersymmetry, to the gauginos and higgsinos.
  \end{itemize}
In section~\ref{res} we present our results.

\section{RGE at two loops}\label{RGEs}
In the case of Split Supersymmetry, the RGE have been computed up to two-loop order in ref.~\cite{GR} for the gauge couplings and
in ref.~\cite{Binger} for the Higgs quartic coupling. We have recalculated these RGE and collected them in  appendix~\ref{RGE}. 

Of course the full set of RGE, and not only the RGE for the Higgs quartic coupling, is needed to get a consistent result for the Higgs mass prediction at a given order in perturbation theory.
Here we have computed the RGE for the third-generation Yukawa couplings $g_t,g_b,g_\tau$ and for the `split' Yukawa-like gaugino couplings
$\tilde{g}_{\rm 1u}$, $\tilde{g}_{\rm 1d}$, $\tilde{g}_{\rm 2u}$, $\tilde{g}_{\rm 2d}$, defined by the interactions
\beq
\Lag = -\frac{H^\dagger}{\sqrt{2}}\left( \tilde{g}_{\rm 2u}\sigma^a {\tilde W}^a+ \tilde{g}_{\rm 1u}{\tilde B}\right) {\tilde H}_u +
\frac{H^T i \sigma_2}{\sqrt{2}}\left( \tilde{g}_{\rm 2d}\sigma^a {\tilde W}^a- \tilde{g}_{\rm 1d}{\tilde B}\right) {\tilde H}_d + {\rm h.c.},
\eeq
where $\sigma^a$ are the Pauli matrices.
These couplings satisfy the matching conditions at the scale $\tm$:
\bea
 \tilde{g}_{\rm 2u}(\tm ) = g_2(\tm ) \sin \beta ,&&~~~~  \tilde{g}_{\rm 1u}(\tm ) = \sqrt{3/5}\, g_1(\tm ) \sin \beta \\
 \tilde{g}_{\rm 2d}(\tm ) = g_2(\tm ) \cos \beta ,&&~~~~  \tilde{g}_{\rm 1d}(\tm ) = \sqrt{3/5}\, g_1(\tm ) \cos \beta .
 \eea
 
The complete set of two-loop RGE for the coupling constants of Split Supersymmetry using $\overline{\rm MS}$ regularization
 is presented in appendix~\ref{RGE}. We find that the dominant 2 loop correction to the predicted
Higgs mass comes from the RGE for the top quark Yukawa coupling, \eq{top2loop}, rather than from the RGE for the quartic Higgs coupling, \eq{lambda2loop}.

\section{Matching at the high scale}\label{hi}

Both for High-Scale and Split Supersymmetry, the potential for the Higgs doublet $H$ below the scale $\tm$ is given by the Standard Model expression
\beq
V(H) =\frac \lambda 2 \left( H^\dagger H -v^2\right)^2,
\eeq
where $v=174$~GeV and the physical Higgs mass is $m_h^2=2\lambda v^2$. The tree-level matching condition with the full supersymmetric theory implies that the quartic coupling 
\beq
\lambda (\tm )= \frac14\left[g_2^2(\tm)+\frac35 g_1^{2}(\tm)\right] \cos^22\beta 
\label{match2}
\eeq 
must be within the range
\beq 
0<\lambda (\tm )< \frac14\left[g_2^2(\tm)+\frac35 g_1^{2}(\tm)\right],
\label{match}
\eeq
where $\beta$ is the rotation angle that defines the light Higgs field, and $g_1$, $g_2$ are weak gauge coupling constants.

At the next-to-leading order, we need to include threshold corrections, and the Higgs quartic coupling at the matching condition becomes $\lambda + \delta\lambda$.
The threshold corrections depend on the masses of the heavy sparticles,
described by the usual parameters $m_L, m_E$ (slepton masses), $m_U,m_D,m_Q$ (squark masses), 
$m_A$ (heavy Higgs mass parameter) and, in the case of High-Scale Supersymmetry, by $M_1,M_2,M_3,\mu$ (gaugino and higgsino masses).
%
We find\footnote{This expression differs from the one in the previous version of this paper as follows.
We added wave function renormalisation terms 
following~\cite{Slavich14} and the definition of $\tan\beta$ at 1 loop precision adopted there.
Furthermore we corrected errors in the heavy-Higgs contributions and re-expressed the result in terms
of the $\overline{\rm MS}$  gauge couplings $g_1, g_2$ of the low-energy non-supersymmetric theory, as in~\cite{Slavich14}. }
\begin{eqnarray}
(4\pi)^2\delta \lambda(\tm) &=&-\bigg[\frac{9}{100}g_1^4+\frac{3}{10}g_1^2g_2^2+\nonumber
(\frac{3}{4}-\frac{\cos^22\beta}{6})g_2^4\bigg]+\\
&&+ 3 g_t^2\bigg[g_t^2 +\frac{1}{10}(5g_2^2-g_1^2)\cos2\beta\bigg]\ln\frac{m_Q^2}{\tm^2}+
3 g_t^2\bigg[g_t^2 +\frac{2}{5}g_1^2\cos2\beta\bigg]\ln\frac{m_U^2}{\tm^2}+\nonumber
\\
&&+3g_t^4 \left[ 
2X_t F(\frac{m_Q}{m_U}) -\frac{X_t^2}{6} G(\frac{m_Q}{m_U}) \right]
+\frac{3}{4} X_t  g_t^2 \bigg[ g_2^2 H_2(\frac{m_Q}{m_U}) +\frac 35 g_1^2 H_1(\frac{m_Q}{m_U})\bigg] \cos2\beta+
\nonumber \\
&&
-\frac{g_t^2}{4}   X_t \cos^2 2\beta 
\left( \frac35 g_1^2 +g_2^2 \right)  H(\frac{m_Q}{m_U}) -3\frac{(5g_2^2 + 3 g_1^2)^2}{400}\sin^24\beta +\nonumber
\\ \nonumber
&&+\frac{1}{4800}  \bigg[261 g_1^4+630 g_1^2 g_2^2  +1325
 g_2^4  - 4 \cos 4 \beta  \left(9 g_1^4+90 g_1^2 g_2^2+175 g_2^4\right) +
\nonumber \\
&&-9 \cos 8\beta  \left(3 g_1^2+5 g_2^2\right)^2 \bigg] \ln \frac{m_A^2}{\tm^2}
+\frac{ \cos^2 2 \beta}{100}
\bigg[18 g_1^4 \ln \frac{m_{E}^2}{\tm^2}+ \nonumber \\
&& +24 g_1^4 \ln \frac{m_{U}^2}{\tm^2} 
+6 g_1^4 \ln \frac{m_{D}^2}{\tm^2} +
3 \left(g_1^4+25 g_2^4\right) 
\ln \frac{m_{Q}^2}{\tm^2} 
 + \left(9 g_1^4+25 g_2^4\right) \ln \frac{m_{L}^2}{\tm^2}   
 \bigg] + \nonumber \\
&&+\tilde\delta_\lambda 
-   
\frac16\,\cos^22\beta \,\left[
  2\,g_2^4\,\ln\frac{M_2^2}{\tm^2}
  +\left(\frac9{25} g_1^4+g_2^4\right)\ln\frac{\mu^2}{\tm^2}\right].
\label{threstop}
\end{eqnarray}
The upper line in eq.~(\ref{threstop}) is the conversion factor from $\overline{\rm DR}$ to $\overline{\rm MS}$ scheme, which modifies
the tree-level  relation of eq.~(\ref{match2}) even in the supersymmetric limit.
The other lines describe the threshold corrections, computed using the $\overline{\rm DR}$ scheme,
from Yukawa contributions of the scalar partners of the top quark (second line);
from stop mixing (3rd and 4th line);
from the heavy Higgses (4th to 6th line);
from the gauge contribution of squarks and leptons (6th and 7th line). Finally, the last line
describes the contributions from gauginos and higgsinos, 
which must be included at the high scale only in the case of High-Scale Supersymmetry. 
For Split Supersymmetry $\tilde \delta_\lambda$  (whose
explicit expression of is  given later in~\eq{eq:thrrr2}) is instead present at the weak scale,
and the other Higgsino/gaugino terms are accounted by RGE running between the weak scale and the SUSY scale.

Here $g_t=m_t/v$ is the top Yukawa coupling, $X_t ={(A_t-\mu \cot \beta)^2}/m_Qm_U$ is the stop mixing parameter,
and
\beq
F(x)=\frac{2x\ln x}{x^2-1}~~~~~~~G(x)=\frac{12x^2\left[ 1-x^2 +(1+x^2)\ln x \right]}{(x^2-1)^3}\qquad
H(x) = \frac{3 x (1 - x^4 + 2 x^2 \ln x^2)}{(1 - x^2)^3}.
\eeq
\beq
H_1(x) = \frac{2x [ 5(1-x^2)+2(1+4x^2)\ln x]}{3(x^2-1)^2}~~~~~~
H_2(x) = \frac{2x ( x^2-1-2\ln x)}{(x^2-1)^2},
\eeq
where the functions are normalized such that $F(1)=G(1)=H(1)=H_{1,2}(1)=1$.

The $g_t^4$ correction is maximized for $X_t \simeq 6$ with a mild dependence on the ratio of the two stop masses (which, for simplicity, is set equal to 1). So, for nearly degenerate squarks, the largest threshold correction comes from the stop sector and
\beq
\delta \lambda_{\rm max} (\tm )=\frac{9g_t^4}{8\pi^2}.
\label{delmax}
\eeq
In the case of Split Supersymmetry the threshold correction in \eq{delmax} is completely negligible because the soft parameters $A_t$ and $\mu$ are both of the order of the weak scale, and thus $X_t = \order{m_W^2/\tm^2}$. We can then take $\delta \lambda_{\rm max} (\tm )=0$, rather than the value given in \eq{delmax}.   
Extra contributions to the matching conditions of the Higgs quartic
coupling in Split Supersymmetry were considered in ref.~\cite{Mah}.

As shown in ref.~\cite{nomura}, the impact on the Higgs mass of the threshold correction in \eq{delmax} is fairly negligible for values of $\tm$ around the GUT scale. This happens because of two effects: $g_t$ at the GUT scale is about half its low-energy value and the renormalization flow of $\lambda$ has a focusing effect: its value at the weak scale is dominated by RGE corrections not much above the weak scale, where $g_t$ is larger.
Nevertheless, the effect of $\delta \lambda$ is important for our analysis because here we are interested in studying the Higgs mass prediction for any value of the supersymmetry-breaking scale, and not only for $\tm$ around the GUT scale. 
For instance we find that the correction to $\lambda(\tm)$ in \eq{delmax}
increases the Higgs mass by only $0.5\GeV$ for $\tm =10^{15}\GeV$, but the effect grows to $6\GeV$ when $\tm =10^5 \GeV$.


\medskip


In the case of Split Supersymmetry we also need the
gaugino couplings renormalized in the $\overline{\rm MS}$ scheme at the $\tm$ scale.  In the limit
of degenerate scalars they are~\cite{Slavich14}:\footnote{We correct the equations below,
with respect to the previous version of this paper, where we had adopted the results of~\cite{LG}.}
\begin{eqnarray}
\frac{\tilde{g}_{\rm 2u}}{g_2\sin\beta} &=& 1+ \frac{1}{(4\pi)^2}\bigg[-g_2^2(\frac{2}{3}+\frac{11}{16}\cos^2\beta)+\frac{3g_1^2}{80}(-2+7\cos^2\beta)+\frac{9g_t^2}{4\sin^2\beta}\nonumber
\bigg],\\
\frac{\tilde{g}_{\rm 2d}}{g_2\cos\beta} &=& 1+ \frac{1}{(4\pi)^2}\bigg[-g_2^2(\frac{2}{3}+\frac{11}{16}\sin^2\beta)+\frac{3g_1^2}{80}(-2+7\sin^2\beta)\nonumber
\bigg],\\
\frac{\tilde{g}_{\rm 1u}}{\sqrt{3/5}g_1\sin\beta} &=& 1+ \frac{1}{(4\pi)^2}\bigg[\frac{3g_2^2}{16}(-2+7\cos^2\beta)+\frac{3g_1^2}{80}(-44+7\cos^2\beta)+\nonumber
\frac{9g_t^2}{4\sin^2\beta}\bigg],\\
\frac{\tilde{g}_{\rm 1d}}{\sqrt{3/5}g_1\cos\beta} &=& 1+ \frac{1}{(4\pi)^2}\bigg[\frac{3g_2^2}{16}(-2+7\sin^2\beta)+\frac{3g_1^2}{80}(-44+7\sin^2\beta)\bigg].
\end{eqnarray}

\subsection{Non-minimal contributions to $\lambda $ at tree level}

It is important to emphasize that the Higgs mass prediction based on the supersymmetric matching condition in \eq{match2} relies on strong assumptions on the behavior of the theory in the far ultraviolet. First, below the scale $\tm$ no new degrees of freedom must be present in order to preserve the renormalization group flow as predicted by the Standard Model (for High-Scale Supersymmetry) or by Split Supersymmetry. Second, any new heavy particle at the scale $\tm$ should not have large couplings to the Higgs superfields and modify \eq{match2} by sizable tree-level or loop effect.

An example of an effect of the first kind is given by right-handed neutrinos with mass $M$, which affect the running between $M$ and $\tm$. This will be discussed in sect.~\ref{concl}.
Concerning the second issue,
let us consider for example a heavy singlet superfield $N$ coupled to the two Higgs doublets with superpotential and supersymmetry breaking interactions given by
\beq
\mathscr{W}=\lambda_N N H_u H_d +\frac{M}{2} N^2,
\eeq
\beq
-\Lag_{\rm soft} = m^2 |N|^2 + \left( A\lambda_N  N H_u H_d +\frac{BM}{2} N^2+ {\rm h.c.}\right) ,
\eeq
where all mass parameters $M$, $m$, $A$, $B$ are of the order of $\tm$. In this case we find that the matching condition in \eq{match} is shifted by an amount
\beq
\delta \lambda = \frac{\lambda_N^2 [(B-2A)M+m^2-A^2]\sin^22\beta}{2(M^2+m^2+BM)}   .
\label{thresing}
\eeq
Potentially this is a large effect that can invalidate our analysis based on the simplest supersymmetric matching condition. Note that the correction in \eq{thresing} can be either positive or negative, and therefore can modify both the upper and lower bounds in \eq{match}. The effect of \eq{thresing} is important especially for moderate values of $\tan\beta$. However, an analogously sizable shift $\delta \lambda$ in the large $\tan\beta$ region can be obtained if the Higgs doublets are coupled to new heavy weak triplet superfields ($T$ and $\bar T$) with a superpotential  
\beq
\mathscr{W}=\lambda_T T H_u H_u +M_T T\bar T .
\eeq
Moreover, important corrections to \eq{match2} are also present if the Higgs doublets are charged under some new gauge forces present at the high scale $\tm$.

\medskip

One could also imagine more unconventional scenarios in which the matching condition of the Higgs quartic at the scale $\tm$ does not respect \eq{match2}. For instance note that, once we assume that supersymmetry is broken at a very high scale, the existence of an R-parity, or of other kinds of matter parities, is no longer a necessary requirement. Indeed, the familiar accidental symmetries of the SM (baryon number, lepton number, lepton flavor...) are automatically present at low energy.

One possibility is that the high-energy theory contains no Higgs superfield and that a single linear combination of the three families of sleptons remains light (as a result of an unnatural fine tuning) playing the role of the low-energy Higgs doublet scalar field. In this case
both the charged lepton and neutrino masses could be generated by the same term in the superpotential
$W=\lambda_{ijk} L_i L_j E_k/2$, with
$\langle \tilde{L}_i\rangle = v_i$.
At tree level the model predicts the following mass matrices for charged leptons and neutrinos
\beq 
m^\ell_{jk} =2 \sum_i v_i \lambda_{ijk},\qquad m_{ij}^\nu= \frac{g_2^2 v_i v_j}{M_2},
\eeq
where $M_2$ is the mass of the superheavy gaugino, which creates a seesaw effect. Other heavy fields and interactions are needed to generate a realistic low-energy mass spectrum. Indeed,
the lepton $L_i$ aligned with $v_i$ remains massless and, at tree level, only its corresponding neutrino gets a mass term. The down-quark mass matrix could be generated by the superpotential $W=\lambda^\prime_{ijk} Q_i D_j L_k$, but other sources are needed to generate the up-quark mass matrix.

Another possibility is matter-Higgs unification in $E_6$.
With supersymmetry at the weak scale, the minimal $E_6$ model involves three generations of $27_i$ for matter and
two pairs of $27\oplus\overline{27}$ for the Higgses.
The latter are no longer necessary if sparticles are heavy enough: the SM can be the low energy limit of 
an $E_6$ theory with just three $27_i$ chiral superfields (that unify all SM fermions and Higgses)
and superpotential
\beq \mathscr{W} =  \lambda_{ijk}27_i 27_j 27_k  .  \eeq
The scalar singlets in
\beq 27_{E_6}=16_{{\rm SO}(10)}\oplus10_{{\rm SO}(10)}\oplus 1_{{\rm SO}(10)}\qquad\hbox{and}\qquad
16_{{\rm SO}(10)} = 10_{{\rm SU}(5)}\oplus\bar 5_{{\rm SU}(5)}\oplus 1_{{\rm SU}(5)}    \eeq
can get vevs, breaking $E_6\to {\rm SO(10)}\to \rm \SU(5)$.
To perform the breaking to the SM group at perturbative 4d level
an additional adjoint 78 superfield is needed. This cannot have any
renormalizable interaction, but it can couple to the $27$'s at non-renormalizable level,
giving SU(5)-breaking fermion masses.
The low-energy Higgs can reside in the 9 different weak doublets of the model (3 for each generation) and be light as a result of an unnatural fine tuning. The
$10-3$ real parameters in the symmetric matrix $\lambda_{ijk}$,
together with the 8 parameters that describe where the light Higgs doublet resides, 
can fit the observed fermion mass matrices. In theories of this kind, the prediction for the Higgs mass is drastically changed.

\medskip

In this paper we will work under the assumption that unknown heavy particles do not strongly couple to the Higgs doublets and that the matching condition at the scale $\tm$ is given by \eq{match2}, with possible corrections coming solely from ordinary supersymmetric particles, as described by \eq{threstop}. It should be noted that effects from unknown heavy particles become irrelevant whenever their supersymmetric mass is much larger than the supersymmetry-breaking mass. This can be observed also in the example of \eq{thresing}, since $\delta \lambda$ rapidly becomes small if $M\gg m,A,B$.

\section{Matching at the weak scale}\label{low}
Consistency of the next-to-leading order calculation requires the inclusion of the one-loop threshold effects at the weak scale.
At one loop order, the pole Higgs and top masses ($m_h$ and  $m_t$) are
related to the Higgs quartic coupling $\lambda(Q)$ and top quark Yukawa coupling $g_t(Q)$
renormalized at the $\overline{\rm MS}$ scale $Q$ as: 
\beq 
m_h^2 = 2v^2 [\lambda(Q)+ \delta_\lambda(Q)+\tilde{\delta}_\lambda(Q)],\qquad
m_t =\frac{g_t(Q)v}{1+\delta_t(Q)+ \tilde{\delta}_t(Q)},
\eeq
where $ v =2^{-3/4}G_{\rm F}^{-1/2}=174.1\GeV$ is extracted from the Fermi constant for muon decay, $G_{\rm F}$.
Here $\delta_\lambda$~\cite{Sirlin} and $\delta_t$~\cite{deltat} are the well-known corrections due to SM particles
\begin{eqnarray}
\delta_\lambda &=& - \frac{\lambda G_{\rm F}M_Z^2}{8\pi^2\sqrt{2}} (\xi F_1 + F_0 + F_3/\xi)\approx 0.0075\lambda \\
\delta_t &=&\delta_t^{\rm QCD} + \delta_t^{\rm EW}\approx -0.0602+0.0013\\
\delta_t^{\rm QCD}(m_t) &=&
-\frac{4}{3\pi}\alpha_3(m_t) -0.92\alpha_3^2(m_t)-2.64\alpha_3^3(m_t) 
   \end{eqnarray}
where  $\xi = m_h^2/M_Z^2$ and the functions $F_i$ are collected in appendix~\ref{soglie}.
The full 2-loop SM corrections have been computed in~\cite{Giardino}.
The numerical values quoted above correspond to $Q=m_t$, $m_h =125\GeV$ and for the present central values:
\beq \label{mtalpha3} m_t = (173.2\pm0.9)\GeV~\cite{topmass},
\qquad \alpha_3 (M_Z) = 0.1184\pm 0.0007~\cite{alpha3}\ .\eeq   
The  corrections $\tilde{\delta}_\lambda$ and $\tilde{\delta}_t$ are due to the supersymmetric fermions at the weak scale. They are present in the case of Split Supersymmetry but should not be not included in the case of High-Scale Supersymmetry. 
These corrections have been computed in ref.~\cite{Binger,Slavich}
in terms of neutralino and chargino mixing matrices.
Here we give the analytic expressions valid in the limit in which the gaugino and higgsino masses are larger than the Higgs mass, $M_1,M_2,\mu\gg m_h$ (actually the expressions  are already accurate 
for $M_1,M_2,\mu\sim m_t$):
\begin{eqnarray}\label{eq:thrrr1}
\tilde{\delta}_t &=& -\frac{\tilde{\beta}_t}{(4\pi)^2}\ln \frac{\mu}{Q}  
-\frac{1}{(4\pi)^2}  \bigg[
\frac{1}{12}(\tilde{g}_{\text{1d}}^2+\tilde{g}_{\text{1u}}^2)  g(r_1)+   \\ \nonumber &&
+\frac{1}{4}(\tilde{g}_{\text{2d}}^2+\tilde{g}_{\text{2u}}^2) g(r_2)+
\frac{1}{6} \tilde{g}_{\text{1d}} \tilde{g}_{\text{1u}} f(r_1)+
\frac{1}{2}\tilde{g}_{\text{2d}} \tilde{g}_{\text{2u}}  f(r_2)
\bigg]
   \end{eqnarray}   
\begin{eqnarray}   \label{eq:thrrr2}
\tilde{\delta}_\lambda &=&\frac{\tilde{\beta}_\lambda}{(4\pi)^2}\ln \frac{\mu}{Q}    \nonumber
+
\frac{1}{(4\pi)^2} \bigg[
-\frac{7}{12}f_1(r_1)   \left(\tilde{g}_{\text{1d}}^4+\tilde{g}_{\text{1u}}^4\right)  
-\frac{9}{4} f_2(r_2) \left(\tilde{g}_{\text{2d}}^4+\tilde{g}_{\text{2u}}^4\right)  +\\  &&  \nonumber
   -\frac{3}{2} f_3(r_1) \tilde{g}_{\text{1d}}^2 \tilde{g}_{\text{1u}}^2
     -\frac{7}{2} f_4(r_2) \tilde{g}_{\text{2d}}^2   \tilde{g}_{\text{2u}}^2
-\frac{8}{3} f_5(r_1,r_2) \tilde{g}_{\text{1d}} \tilde{g}_{\text{1u}} \tilde{g}_{\text{2d}} \tilde{g}_{\text{2u}}
     +\\ \nonumber &&
  -\frac{7}{6} f_6(r_1,r_2) \left(\tilde{g}_{\text{1d}}^2\tilde{g}_{\text{2d}}^2+\tilde{g}_{\text{1u}}^2 \tilde{g}_{\text{2u}}^2\right)
    -\frac{1}{6} f_7(r_1,r_2) \left(\tilde{g}_{\text{1d}}^2
   \tilde{g}_{\text{2u}}^2+\tilde{g}_{\text{1u}}^2 \tilde{g}_{\text{2d}}^2\right)
   +\\  &&
-\frac{4}{3} f_8(r_1,r_2)
   \left(\tilde{g}_{\text{1d}} \tilde{g}_{\text{2u}}+\tilde{g}_{\text{1u}} \tilde{g}_{\text{2d}}\right) \left(\tilde{g}_{\text{1d}}
   \tilde{g}_{\text{2d}}+\tilde{g}_{\text{1u}} \tilde{g}_{\text{2u}}\right)
      +\\  &&   \nonumber
   +\frac{2}{3} f\left(r_1\right) \tilde{g}_{\text{1d}}
   \tilde{g}_{\text{1u}} \left[\lambda -2 \left(\tilde{g}_{\text{1d}}^2+\tilde{g}_{\text{1u}}^2\right)\right]
      +2 f\left(r_2\right) \tilde{g}_{\text{2d}} \tilde{g}_{\text{2u}}
   \left[\lambda -2 \left(\tilde{g}_{\text{2d}}^2+\tilde{g}_{\text{2u}}^2\right)\right]
   +\\  &&  \nonumber
 +  \frac{1}{3} g\left(r_1\right) \lambda 
   \left(\tilde{g}_{\text{1d}}^2+\tilde{g}_{\text{1u}}^2\right)
+ g\left(r_2\right)  \, \lambda  \left(\tilde{g}_{\text{2d}}^2+\tilde{g}_{\text{2u}}^2\right)   
   \bigg].
   \end{eqnarray}
%
%
%
Here $r_1=M_1/\mu$, $r_2=M_2/\mu$, 
and the functions $f,g,f_i$ are defined in appendix~\ref{soglie}. All these function are normalized such that they are equal to 1 when their arguments equal 1.
The first terms in eqs.~(\ref{eq:thrrr1}) and (\ref{eq:thrrr2}) are the scale-dependent contributions and their coefficients are
\begin{eqnsystem}{sys:betatilde}
\tilde{\beta}_t  &=&(\tilde{g}_{\text{1d}}^2+\tilde{g}_{\text{1u}}^2+3
   \tilde{g}_{\text{2d}}^2+3 \tilde{g}_{\text{2u}}^2)/2\\
   \tilde{\beta}_\lambda &=& 2\lambda \left(\tilde{g}_{\text{1d}}^2+\tilde{g}_{\text{1u}}^2+3 \tilde{g}_{\text{2d}}^2+3 \tilde{g}_{\text{2u}}^2\right)
   -\tilde{g}_{\text{1d}}^4
-\tilde{g}_{\text{1u}}^4-5 \tilde{g}_{\text{2d}}^4-5 \tilde{g}_{\text{2u}}^4
\\ \nonumber &&
-4 \tilde{g}_{\text{1d}} \tilde{g}_{\text{1u}} \tilde{g}_{\text{2d}} \tilde{g}_{\text{2u}}-2 \left(\tilde{g}_{\text{1d}}^2+\tilde{g}_{\text{2u}}^2\right) \left(\tilde{g}_{\text{1u}}^2+\tilde{g}_{\text{2d}}^2\right) .
\end{eqnsystem}
These correspond to the Split Supersymmetry contribution to the RGE of $g_t$ and $\lambda$ respectively, see eq.~(\ref{sys:RGE1}),
such that the dependence on the renormalization scale $Q$ cancels out at leading one loop order.

The threshold corrections to the gauge couplings $g$ are well known. If one wants to
apply the RGE of Split Supersymmetry from a low-energy matching scale $Q$ up to $\tilde m$
one needs to employ
\begin{eqnsystem}{sys:g} 
g_1(Q)|_{\rm Split} &=& g_1(Q)|_{\rm SM}  -\frac{g_1^3}{(4\pi)^2}\frac{2}{5}\ln \frac{\mu}{Q}\\
g_2(Q)|_{\rm Split} &=& g_2(Q)|_{\rm SM}  -\frac{g_2^3}{(4\pi)^2}\bigg(\frac{4}{3}\ln \frac{M_2}{Q}+\frac{2}{3}\ln\frac{\mu}{Q}\bigg)\\
g_3(Q)|_{\rm Split} &=& g_3(Q)|_{\rm SM}  -\frac{g_3^3}{(4\pi)^2}2\ln \frac{M_3}{Q}\   .  \label{eq:g3}
\end{eqnsystem}
In the following we will fix the following unified spectrum of gauginos and higgsinos
\beq M_1  = m_t, \qquad M_2 = \mu = 2 M_1,\qquad M_3 = 6.4 M_1.\label{eq:Mi}\eeq


\begin{figure}[t]
$$\includegraphics[width=0.45\textwidth]{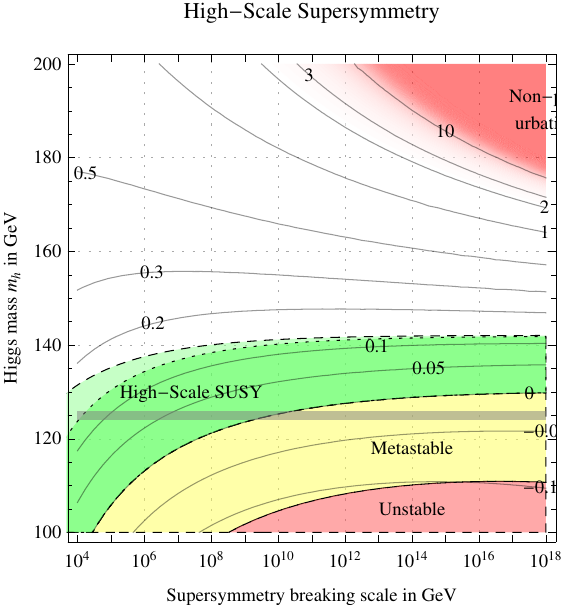}\qquad \includegraphics[width=0.45\textwidth]{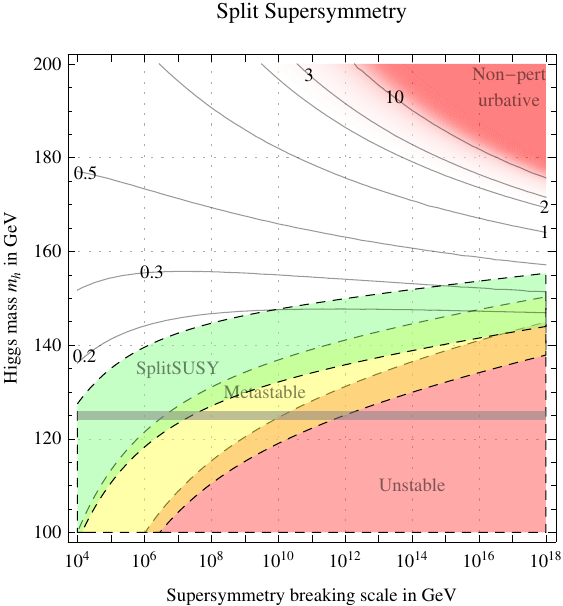}$$
\caption{\em Contour plot of the Higgs quartic coupling renormalized at the supersymmetry breaking scale $\tm$.
The  regions marked as ``metastable'' (yellow) and ``unstable'' (red)
correspond to $\lambda<0$;
the green band shows the range of the Higgs mass allowed by the supersymmetric matching condition for the Higgs quartic coupling, in the case of 
High-Scale Supersymmetry (left panel; the dashed and dotted curves correspond to the cases of
maximal and minimal stop threshold corrections)
and
Split Supersymmetry (right panel, dashed curves;
double contour-lines and partially overlapped regions are due to the variation with $\tan\beta$
of the gaugino couplings).
The values of $\alpha_3$ and $m_t$ are fixed to their central values, see eq.~(\ref{mtalpha3}),
and the horizontal band $124\GeV<m_h<126\GeV$ shows the experimentally favored range.
\label{fig:models}}
\end{figure}

\begin{figure}[t]
$$\includegraphics[width=0.45\textwidth]{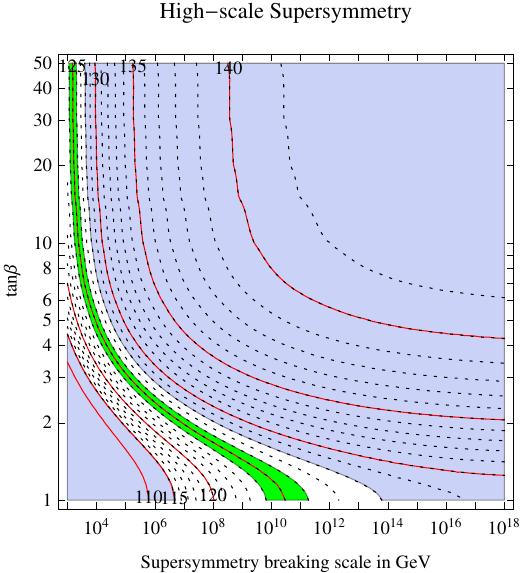}\qquad 
\includegraphics[width=0.45\textwidth]{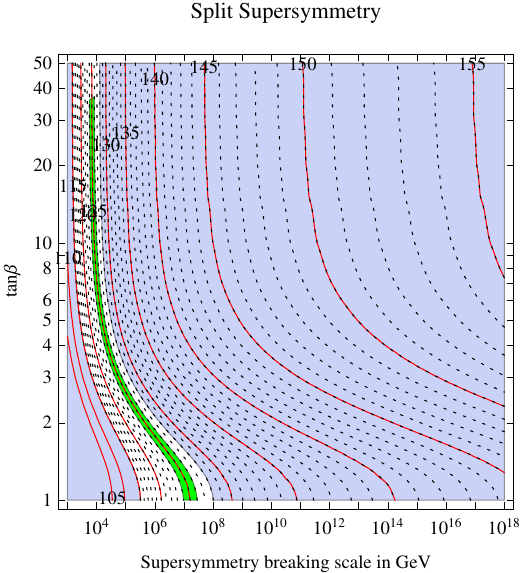}$$
\caption{\em Prediction for the Higgs mass $m_h$ at two loops in High-Scale Supersymmetry (left panel) and Split Supersymmetry (right panel) 
as a function of the supersymmetry breaking scale $\tilde m$ and $\tan\beta$ for the central values of $\alpha_3$ and $m_t$. In the case of Split Supersymmetry we have chosen the light sparticle spectrum of \eq{eq:Mi};
in the case of High Scale Supersymmetry we assumed maximal stop mixing.
Excluded values $m_h<115\GeV$ and $m_h>128\GeV$ are shaded in gray;
the favorite range $124\GeV<m_h < 126\GeV$ is shaded in green.
\label{fig:mh}}
\end{figure}

\begin{figure}[t]
$$
\includegraphics[width=0.7\textwidth]{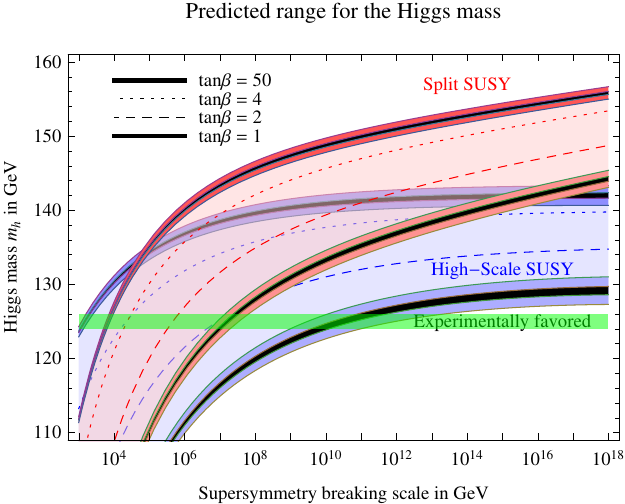}$$
\caption{\em Next-to-leading order prediction for the Higgs mass $m_h$ in High-Scale Supersymmetry (blue, lower) and Split Supersymmetry (red, upper) 
for $\tan\beta=\{1,2,4,50\}$. The thickness of the lower boundary at $\tan\beta=1$ and of the upper boundary at $\tan\beta=50$ 
shows the uncertainty due to the present $1\sigma$ error on $\alpha_3$ (black band) and on the top mass (larger colored band).
\label{fig:mhrange}}
\end{figure}

\begin{figure}[t]
$$\includegraphics[width=0.7\textwidth]{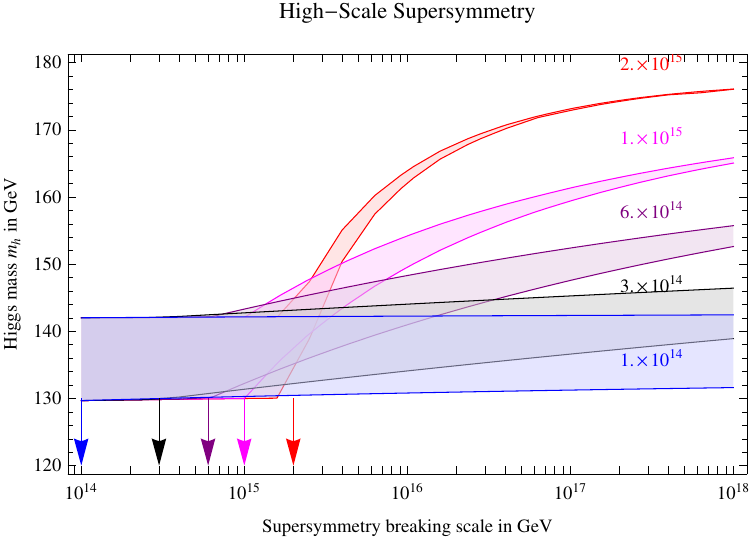}$$
\caption{\em The impact of neutrino Yukawa couplings
on the predicted range for the Higgs mass in High-Scale Supersymmetry assuming best-fit values
for $m_t$ and $\alpha_3$ and varying $\tan\beta$.
Each band corresponds to a different  value of the right-handed neutrino mass,  
as indicated in the figure. The arrows show the points where $\tm =M$, below which the effect disappears.
\label{fig:mnu}}
\end{figure}

\begin{figure}
$$\includegraphics[width=0.45\textwidth]{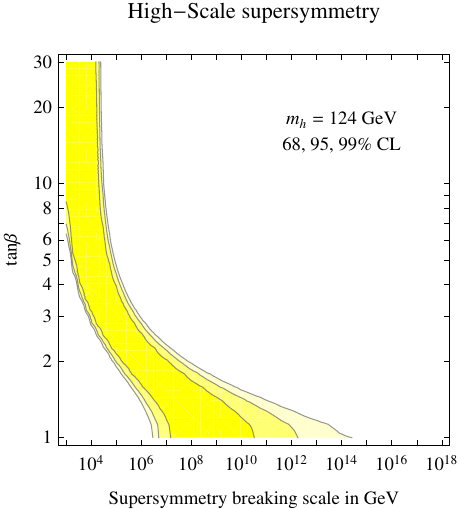}\qquad\includegraphics[width=0.45\textwidth]{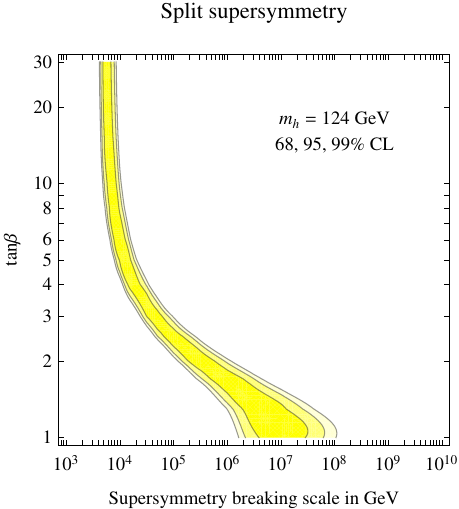}$$
$$\includegraphics[width=0.45\textwidth]{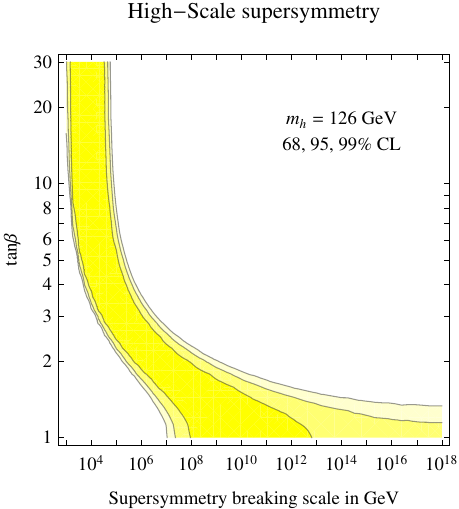}\qquad\includegraphics[width=0.45\textwidth]{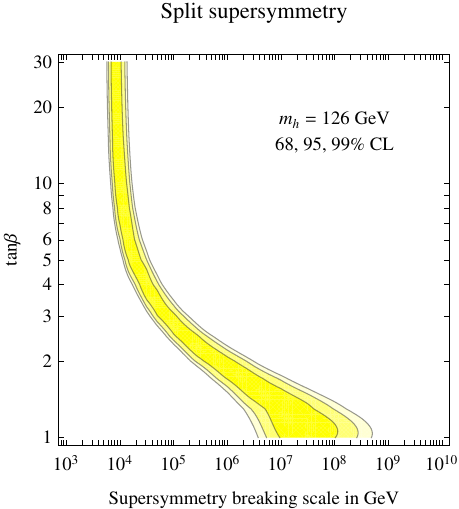}$$
\caption{\em Assuming the existence of supersymmetry we compute, as function of $\tan\beta$,
the preferred value of the SUSY scale $\tm$
implied by the Higgs mass $m_h=124\GeV$ (upper) and $126\GeV$ (lower) at $68,90,99\%$ C.L.
in the cases of High-Scale Supersymmetry (left, assuming
a degenerate sparticle spectrum at the SUSY breaking scale with arbitrary stop mixing) and Split Supersymmetry (right,
assuming the spectrum of light fermions in \eq{eq:Mi} and a degenerate sparticle spectrum at the SUSY breaking scale).
\label{fig:fixedmh}}
\end{figure}

   \section{Results}\label{concl}\label{res}
   Figure\fig{models} shows the values of the Higgs quartic coupling $\lambda$ renormalized at a given high-energy scale $\tilde m$, as functions
   of $\tilde m$ (horizontal axis) and $m_h$ (vertical axis).
   
The plane is divided in various regions:
\begin{itemize}
\item The yellow middle region marked as ``Metastable''
corresponds to $\lambda <0$ at the high scale, such that the electroweak vacuum is unstable, although its decay rate is smaller than the age of the universe~\cite{vacdecay}.
\item The lower red region marked as ``Unstable'' corresponds to large and negative values of $\lambda$ at the high scale, which trigger an exceedingly fast vacuum decay~\cite{vacdecay}.
\item The upper red region marked as ``Non-perturbative" corresponds to a  large Higgs coupling violating the requirement of perturbativity. This region has already been completely excluded by recent LHC data on Higgs searches.
\end{itemize}
In the case of Split Supersymmetry there is a partial overlap between these regions shown in fig.~\fig{models} because the RGE involve the
gaugino couplings which depend on the unknown parameter $\tan\beta$. Therefore $m_h$ does not uniquely determine the RG trajectory of the Higgs quartic coupling $\lambda$ below $\tm$.

The regions described so far have no connection with the identification of $\tilde m$ with the supersymmetry breaking scale. In this paper we are mostly interested in the last region:
\begin{itemize}
\item The green region covers the range of $m_h$ and $\tm$ allowed by High-Scale Supersymmetry (left panel) and Split Supersymmetry (right panel), as determined by \eq{match}.
In the case of High-Scale Supersymmetry, the boundary is computed both including (dashed line) and ignoring (dotted line) the finite threshold correction of \eq{delmax}.
\end{itemize}

In fig.\fig{mh} we show the predicted Higgs mass $m_h$ as a function of $\tan\beta$ and of the supersymmetry breaking scale $\tilde m$.
Values $m_h>128\GeV$ and $m_h<115\GeV$ have been experimentally excluded and are shaded in gray.
So far in the analysis we assumed the best fit values for $m_t$ and $\alpha_3$, see \eq{mtalpha3},  computed the Split Supersymmetry thresholds at the weak scale assuming \eq{eq:Mi}, and computed the thresholds at the high scale assuming degenerate
sparticles at the scale $\tilde{m}$ and (in the case of High Scale Supersymmetry) maximal stop mixing.

\smallskip

Next we want to study the uncertainty in the Higgs mass prediction due to the errors in $m_t$ and $\alpha_3$.
In fig.\fig{mhrange} we show the allowed ranges for $m_h$ as functions of $\tilde{m}$, taking into account experimental uncertainties:
the boundaries at $\tan\beta=1$ and at large $\tan\beta$ are computed varying $\alpha_3$ (black bands) and $m_t$ (colored bands) by $\pm 1\sigma$. The largest uncertainty comes from the measurement of $m_t$ and corresponds to a 1-$\sigma$ error in $m_h$ of about 1--1.5 GeV,
depending on $\tilde m$ and $\tan\beta$.
We assume maximal stop mixing in the case of High Scale Supersymmetry at large $\tan\beta$, and zero
stop mixing otherwise.
Of course, the unknown sparticle mass spectra provide extra uncontrollable uncertainties.


\medskip

Finally we study the effect of the couplings needed to generate neutrino masses.
We assume type-I see-saw and fix  the largest right-handed neutrino Yukawa coupling to its ``minimal'' value,
$g_\nu = \sqrt{m_{\rm atm} M}/v$, where $M$ is the right-handed neutrino mass and
$m_{\rm atm} \approx 0.06\,{\rm eV}$ is the light neutrino mass renormalized at $M$.
Taking into account its RGE effects at two loops, we find that, for $\tilde m> M$,  the predicted Higgs mass in High-Scale Supersymmetry increases as shown in fig.\fig{mnu}.
The effect is roughly equivalent to the following correction to the high-energy
matching condition:
\beq \delta \lambda (\tilde m)\simeq \frac{M m_\nu}{4\pi^2 v^2} \ln \frac{\tilde m}{M}\qquad \hbox{for $\tilde{m}>M$}\label{eq:nu}\eeq
which is irrelevant if $M \circa{<} 10^{14}\GeV$.

\subsection{Implications of present Higgs searches at the LHC}
Recent data from ATLAS and CMS provide a 99\% CL upper bound on the SM Higgs mass of 128 GeV and a hint in favor of a Higgs mass 
in the $124-126\GeV$ range~\cite{mhexp}.
The main implications for the scale of supersymmetry breaking can be read from fig.\fig{mhrange} and are more precisely
studied in fig.\fig{fixedmh}, where we perform a fit taking into account the experimental uncertainties on
the top mass and the strong coupling.

The scale of Split Supersymmetry is constrained to be below a few $10^8\GeV$. This implies a significant upper bound on the gluino lifetime~\cite{slavich}
\beq
\tau_{\tilde g} \simeq \left( \frac{\rm TeV}{M_{3}} \right)^5 \left( \frac{\tm}{10^8~{\rm GeV}}\right)^4 4\times 10^{-4} ~{\rm s}.
\eeq
As the value of $\tan\beta$ increases, the bound on $\tm$ becomes rapidly much tighter, see fig.\fig{fixedmh}. For instance, for $\tan\beta >10$, the scale of Split Supersymmetry must be below about $10^4$ GeV and the gluino lifetime must be less than $4\times 10^{-20}(M_{3}/ {\rm TeV})^{-5} ~{\rm s}$.

In the case of High-Scale Supersymmetry, the absolute upper bound on the scale of supersymmetry breaking strongly depends on precise determinations of the SM parameters, especially $m_h$ and $m_t$. Supersymmetry at the weak scale can reproduce the Higgs mass favored by the preliminary ATLAS and CMS analyses provided that
 $\tan\beta$ is large, that stops are in hundreds of GeV to TeV range and are strongly mixed. On the other hand,
supersymmetry at the Planck scale
can  reproduce the  Higgs mass too, provided that $m_h$ lies in the upper part of the favored range, that
$\tan\beta$ is very close to 1, that the stop mixing parameter $X_t$ is negligible, and that the
 top mass is about $2$ standard deviations below its best-fit value
(possible with a strong coupling somewhat above its central value). However, for moderate or large values on $\tan\beta$, the scale of supersymmetry breaking is severely constrained also in the case of High-Scale Supersymmetry, as shown in fig.\fig{fixedmh}.




\medskip

   \section{Conclusions}
In this paper we computed the range of Higgs masses determined by matching the quartic Higgs coupling to its supersymmetric value at a very large energy scale $\tilde m$. We assumed that the matching condition is
precisely given by \eq{match2}, with the possible addition of the correction in \eq{delmax}. This assumption relies on the absence of new large Higgs couplings at the high-energy scale and it can be violated in models with additional singlet or weak triplet chiral superfields or in models where the SM fermions reside in the same supermultiplet as the Higgs boson, realizing a matter-Higgs unification.

We have considered two scenarios: 1)  the  particle content below the scale $\tm$ is described by the SM (High-Scale Supersymmetry) or 
2) the particle content below $\tm$ is described by the SM plus the
supersymmetric fermions needed for dark matter and gauge coupling unification (Split Supersymmetry). In the latter case we presented a new full next-to-leading order analysis, computing the complete RGE at two loops and the one-loop threshold corrections. These effects reduce
the predicted Higgs mass by a few GeV with respect to the one-loop result. 

It is interesting that the measurement of the Higgs mass can give information on the possible presence of supersymmetry even at energy scales too high to have any chance to be directly tested at colliders. 
We studied the implications of a Higgs mass in the range $124-126 \GeV$ recently favored by the ATLAS and CMS experiments.
Our results are summarized in fig.\fig{fixedmh}. We can already infer that the scale of Split Supersymmetry must be $ \tm \circa{<}10^8\GeV$ (unless sparticles threshold corrections are large), while information on High-Scale Supersymmetry is not yet conclusive.  Improved measurements of $m_h$ and $m_t$, together with higher-order
computations of the weak scale thresholds, will be crucial to extract important information on the scale of supersymmetry breaking.

\paragraph{Acknowledgements} We are grateful to P. Slavich for his precious suggestions and for comparing results. We thank A. Djouadi, D. Feldman, P. Giardino, L. Hall, P. Lodone, A. Romanino,  and J. Wells  for useful discussions. 
This work was supported by the ESF grant MTT8 and by SF0690030s09 project.

\appendix
\small
%
%
%

\section{Renormalization Group Equations}\label{RGE}

The RGE up to two-loop order have been presented in ref.~\cite{MV} for a generic gauge field theory.
We write the generic Lagrangian  in terms of vectors $V_A$, real scalars $\phi^a$, and Weyl fermions $\psi_i$ as:
\beq\Lag = -\frac{1}{4}(F_{\mu\nu}^A)^2 + \frac{(D_\mu \phi_a)^2}{2} + i\bar\psi_i \slashed{D} \psi_i-
\frac{1}{2}( Y^a_{ij} \psi_i \psi_j \phi_a+\hbox{h.c.}) -\frac{\lambda_{abcd}}{4!}\phi_a\phi_b\phi_c\phi_d .
\eeq
The gauge covariant derivatives are 
\beq D_\mu \phi_a = \partial_\mu \phi_a - i\theta^A_{ab} V_\mu^A\phi_b,\qquad
D_\mu\psi_i =\partial_\mu \psi_i -i t^A_{ij}V_\mu^A \psi_j .
\eeq
In our notation the generators $\theta^A$ (for scalars) and $t^A$ (for fermions) contain the coupling constant, {\it e.g.}\
$t^A = g_2 \sigma^A/2$ for a SU(2) doublet.
Note that the Yukawa and quartic couplings are written in components and satisfy gauge invariance relations such as
$Y^a \cdot t^A + t^{AT} \cdot Y^a+Y^b\theta^A_{ba}=0$.

Using these results we produced a code that, for any given generic Lagrangian, derives RGE equations up to two-loop order.
We corrected a few problems present in ref.~\cite{MV}, already described in~\cite{luo}.
Unlike~\cite{MV,luo}, we define $C^{AB} = f^{ACD}f^{BCD}$ and
\beq \begin{array}{ll}
C_2^{ab}(S) = (\theta^A \cdot \theta^A)_{ab} &
S_2^{AB}(S) = \Tr (\theta^A\cdot\theta^B)\\
C_2^{ij}(F) = (t^A \cdot t^A)_{ij} &
S_2^{AB}(F) = \Tr (t^A\cdot t^B) .
\end{array}\eeq
These objects satisfy well-known relations demanded by group theory that were used in~\cite{MV,luo} to write results
in terms of group invariants, assuming a 
simple Lie group. 
Avoiding  such simplifications, 
we obtain a unique set of compact generic formul\ae{} not restricted to simple Lie groups.
For example, in the SM, $V^A$ describes the $1+3+8$ vectors;  one generation of fermions is described by 15 $\psi_i$
and the Higgs doublet $H$ is described as a 4 component scalar $\phi_a$.

We write the RGE for each coupling $g_i$ present in the theory, in the $\overline{\rm MS}$ scheme, as
\beq \frac{dg_i}{d\ln Q} = \frac{\beta_1(g_i)}{(4\pi)^2} +\frac{\beta_2(g_i)}{(4\pi)^4} .\eeq
Specializing to the case of Split Supersymmetry, the relevant coupling constants $g_i$ include the gauge couplings ($g_1,g_2,g_3$), the third-generation Yukawa couplings ($g_t,g_b,g_\tau$), the gaugino couplings ($\tilde{g}_{\text{1d}},\tilde{g}_{\text{1u}}$, $\tilde{g}_{\text{2d}},\tilde{g}_{\text{2u}}$), and the Higgs quartic ($\lambda$).

At one loop we recover the results of~\cite{GR} for the $\beta$ functions of Split Supersymmetry, which are given by
\beq\beta_1 (g_1) =\frac{9}{2}g_1^3,\qquad\beta_1(g_2) = -\frac{7}{6} g_2^3,\qquad\beta_1(g_3) =-5 g_3^3\eeq
\begin{eqnsystem}{sys:RGE1}
\beta _1\left(g_t\right) &=& g_t \left( \tilde\beta_t+\frac{3}{2} g_b^2 +g_{\tau }^2 +\frac{9}{2} g_t^2-\frac{17}{20} g_1^2 -\frac{9}{4} g_2^2 -8 g_3^2 \right) \\
\beta _1\left(g_b\right) &=&g_b\left(  \tilde\beta_t+\frac{3}{2}  g_t^2+ g_{\tau }^2+\frac{9}{2} g_b^2-\frac{1}{4} g_1^2 -\frac{9}{4} g_2^2 -8 g_3^2 \right) \\
\beta _1\left(g_{\tau }\right) &=&g_\tau \left( \tilde\beta_t +3 g_b^2 +3 g_t^2 +\frac{5}{2} g_{\tau }^2-\frac{9}{4} g_1^2 -\frac{9}{4} g_2^2 \right) \\
\beta _1\left(\tilde{g}_{\text{2u}}\right) &=& \tilde{g}_{\text{1d}} \tilde{g}_{\text{1u}} \tilde{g}_{\text{2d}}+\tilde{g}_{\text{2u}} \left(\frac{3}{4} \tilde{g}_{\text{1u}}^2 +\frac{1}{2} \tilde{g}_{\text{1d}}^2+\frac{11}{4} \tilde{g}_{\text{2u}}^2+\tilde{g}_{\text{2d}}^2 +3 g_b^2 +3 g_t^2 +g_{\tau }^2-\frac{9}{20} g_1^2-\frac{33}{4} g_2^2 \right)  \\
\beta _1\left(\tilde{g}_{\text{2d}}\right) &=& \tilde{g}_{\text{1d}} \tilde{g}_{\text{1u}} \tilde{g}_{\text{2u}}+\tilde{g}_{\text{2d}} \left(\frac{3}{4} \tilde{g}_{\text{1d}}^2+\frac{1}{2} \tilde{g}_{\text{1u}}^2 +\frac{11}{4} \tilde{g}_{\text{2d}}^2+\tilde{g}_{\text{2u}}^2+3 g_b^2 +3 g_t^2 +g_{\tau }^2-\frac{9}{20} g_1^2 -\frac{33}{4} g_2^2 \right) \qquad\qquad \\
\beta _1\left(\tilde{g}_{\text{1u}}\right) &=& 3 \tilde{g}_{\text{1d}} \tilde{g}_{\text{2d}} \tilde{g}_{\text{2u}}
+\tilde{g}_{\text{1u}} \left(2 \tilde{g}_{\text{1d}}^2+\frac{5}{4} \tilde{g}_{\text{1u}}^2+
\frac{3}{2} \tilde{g}_{\text{2d}}^2+\frac{9}{4}  \tilde{g}_{\text{2u}}^2+3 g_b^2 +3 g_t^2 +g_{\tau }^2-\frac{9}{20} g_1^2-\frac{9}{4} g_2^2 \right) \\
\beta _1\left(\tilde{g}_{\text{1d}}\right) &=& 3 \tilde{g}_{\text{1u}} \tilde{g}_{\text{2d}} \tilde{g}_{\text{2u}} 
+\tilde{g}_{\text{1d}} \left(2  \tilde{g}_{\text{1u}}^2+\frac{5}{4} \tilde{g}_{\text{1d}}^2+
\frac{3}{2}  \tilde{g}_{\text{2u}}^2+\frac{9}{4} \tilde{g}_{\text{2d}}^2+3 g_b^2 +3 g_t^2 +g_{\tau }^2-\frac{9}{20} g_1^2-\frac{9}{4} g_2^2 \right) 
\\
\beta _1(\lambda ) &=& 4\lambda  \left( 3 g_t^2+ 3g_b^2+g_{\tau }^2\right) -9\lambda \left( \frac{g_1^2}{5}+g_2^2\right)\!\!   
-4 \left(3g_t^4+3g_b^4+ g_{\tau }^4\right) +
\\ &&+\frac{27 }{100}g_1^4+\frac{9}{10} g_2^2 g_1^2+\frac{9 }{4}g_2^4+12 \lambda ^2+\tilde{\beta}_\lambda \nonumber
   \end{eqnsystem}
   where $\tilde\beta_t$ and $\tilde\beta_\lambda$ have been defined in eq.~(\ref{sys:betatilde}).

At two loops, we have the following $\beta$ functions for the gauge couplings~\cite{GR}:
   \begin{eqnarray}
\beta _2(g_1) &=& g_1^3 \bigg[\!\!
-\frac{3}{20} (\tilde{g}_{\text{1d}}^2+\tilde{g}_{\text{1u}}^2)-\frac{9}{20} (\tilde{g}_{\text{2d}}^2+\tilde{g}_{\text{2u}}^2)-\frac{g_b^2}{2}-\frac{17 g_t^2}{10}-\frac{3 g_{\tau }^2}{2}+\frac{104 g_1^2}{25}+\frac{18 g_2^2}{5}+\frac{44 g_3^2}{5}\bigg] \\
\beta _2(g_2) &=& g_2^3 \bigg[-\frac{1}{4} (\tilde{g}_{\text{1d}}^2+\tilde{g}_{\text{1u}}^2)-\frac{11}{4} (\tilde{g}_{\text{2d}}^2+\tilde{g}_{\text{2u}}^2)-\frac{1}{2} (3g_b^2+3g_t^2+g_\tau^2)+\frac{6 g_1^2}{5}+\frac{106 g_2^2}{3}+12 g_3^2\bigg] \\
\beta _2(g_3) &=& g_3^3 \bigg[-2 (g_b^2+ g_t^2)+\frac{11 g_1^2}{10}+\frac{9 g_2^2}{2}+22 g_3^2\bigg]
 \end{eqnarray}
The two-loop $\beta$ functions for the third-generation Yukawa couplings are
   \begin{eqnarray}\label{top2loop}
\beta _2(g_t) &=& -12 g_t^5 +g_t \bigg[g_b^2 \bigg(\frac{5 \tilde{g}_{\text{1d}}^2}{8}+\frac{5 \tilde{g}_{\text{1u}}^2}{8}+\frac{15 \tilde{g}_{\text{2d}}^2}{8}+\frac{15 \tilde{g}_{\text{2u}}^2}{8}+\frac{5 g_{\tau }^2}{4}+\frac{7 g_1^2}{80}+\frac{99 g_2^2}{16}+4 g_3^2\bigg)
+\\ \nonumber &&
+g_1^2 \bigg(\frac{3 \tilde{g}_{\text{1d}}^2}{16}+\frac{3 \tilde{g}_{\text{1u}}^2}{16}+\frac{9 \tilde{g}_{\text{2d}}^2}{16}+\frac{9 \tilde{g}_{\text{2u}}^2}{16}-\frac{9 g_2^2}{20}+\frac{19 g_3^2}{15}\bigg)-3 \tilde{g}_{\text{1d}} \tilde{g}_{\text{1u}} \tilde{g}_{\text{2d}} \tilde{g}_{\text{2u}}
+\\ \nonumber &&
+g_2^2 \bigg(\frac{15 \tilde{g}_{\text{1d}}^2}{16}+\frac{15 \tilde{g}_{\text{1u}}^2}{16}+\frac{165 \tilde{g}_{\text{2d}}^2}{16}+\frac{165 \tilde{g}_{\text{2u}}^2}{16}+9 g_3^2\bigg)-\frac{5}{4} \tilde{g}_{\text{1d}}^2 \tilde{g}_{\text{1u}}^2-\frac{9}{8} \tilde{g}_{\text{1d}}^2 \tilde{g}_{\text{2d}}^2-\frac{9 \tilde{g}_{\text{1d}}^4}{16}
+\\ \nonumber &&
-\frac{9}{8} \tilde{g}_{\text{1u}}^2 \tilde{g}_{\text{2u}}^2-\frac{9 \tilde{g}_{\text{1u}}^4}{16}-\frac{3}{4} \tilde{g}_{\text{2d}}^2 \tilde{g}_{\text{2u}}^2-\frac{45 \tilde{g}_{\text{2d}}^4}{16}-\frac{45 \tilde{g}_{\text{2u}}^4}{16}-\frac{g_b^4}{4}-\frac{9 g_{\tau }^4}{4}
+\\ \nonumber &&
+\bigg(\frac{15 g_1^2}{8}+\frac{15 g_2^2}{8}\bigg) g_{\tau }^2+\frac{1303 g_1^4}{600}
-\frac{17 g_2^4}{4}-\frac{284 g_3^4}{3}+\frac{3 \lambda ^2}{2}\bigg] 
+\\ \nonumber &&
+g_t^3 \bigg(-\frac{9 \tilde{g}_{\text{1d}}^2}{8}-\frac{9 \tilde{g}_{\text{1u}}^2}{8}-\frac{27 \tilde{g}_{\text{2d}}^2}{8}-\frac{27 \tilde{g}_{\text{2u}}^2}{8}-\frac{11 g_b^2}{4}-\frac{9 g_{\tau }^2}{4}+\frac{393 g_1^2}{80}+\frac{225 g_2^2}{16}+36 g_3^2-6 \lambda \bigg)
   \end{eqnarray}
  \begin{eqnarray}
\beta _ 2(g_b) &=& -12 g_b^5 +g_b \bigg[g_t^2 \bigg(\frac{5 \tilde{g}_ {\text{1d}}^2}{8}+\frac{5 \tilde{g}_ {\text{1u}}^2}{8}+\frac{15 \tilde{g}_ {\text{2d}}^2}{8}+\frac{15 \tilde{g}_ {\text{2u}}^2}{8}+\frac{5 g_ {\tau }^2}{4}+\frac{91 g_ 1^2}{80}+\frac{99 g_ 2^2}{16}+4 g_ 3^2\bigg)
+\\ \nonumber &&
+g_ 1^2 \bigg(\frac{3 \tilde{g}_ {\text{1d}}^2}{16}+\frac{3 \tilde{g}_ {\text{1u}}^2}{16}+\frac{9 \tilde{g}_ {\text{2d}}^2}{16}+\frac{9 \tilde{g}_ {\text{2u}}^2}{16}-\frac{27 g_ 2^2}{20}+\frac{31 g_ 3^2}{15}\bigg)-3 \tilde{g}_ {\text{1d}} \tilde{g}_ {\text{1u}} \tilde{g}_ {\text{2d}} \tilde{g}_{\text{2u}}
+\\ \nonumber &&
+g_ 2^2 \bigg(\frac{15 \tilde{g}_ {\text{1d}}^2}{16}+\frac{15 \tilde{g}_ {\text{1u}}^2}{16}+\frac{165 \tilde{g}_ {\text{2d}}^2}{16}+\frac{165 \tilde{g}_ {\text{2u}}^2}{16}+9 g_ 3^2\bigg)-\frac{5}{4} \tilde{g}_ {\text{1d}}^2 \tilde{g}_ {\text{1u}}^2-\frac{9}{8} \tilde{g}_ {\text{1d}}^2 \tilde{g}_ {\text{2d}}^2-\frac{9 \tilde{g}_ {\text{1d}}^4}{16}
+\\ \nonumber &&
-\frac{9}{8} \tilde{g}_ {\text{1u}}^2 \tilde{g}_ {\text{2u}}^2-\frac{9 \tilde{g}_ {\text{1u}}^4}{16}
-\frac{3}{4} \tilde{g}_ {\text{2d}}^2 \tilde{g}_ {\text{2u}}^2-\frac{45 \tilde{g}_ {\text{2d}}^4}{16}-\frac{45 \tilde{g}_ {\text{2u}}^4}{16}-\frac{g_t^4}{4}-\frac{9 g_ {\tau }^4}{4}
+\\ \nonumber &&
+\bigg(\frac{15 g_ 1^2}{8}+\frac{15 g_ 2^2}{8}\bigg) g_ {\tau }^2-\frac{131 g_ 1^4}{600}-\frac{17 g_ 2^4}{4}-\frac{284 g_ 3^4}{3}+\frac{3 \lambda ^2}{2}\bigg]+\\ \nonumber &&
+g_b^3 \bigg(-\frac{9 \tilde{g}_ {\text{1d}}^2}{8}-\frac{9 \tilde{g}_ {\text{1u}}^2}{8}-\frac{27 \tilde{g}_ {\text{2d}}^2}{8}-\frac{27 \tilde{g}_ {\text{2u}}^2}{8}-\frac{11 g_t^2}{4}-\frac{9 g_ {\tau }^2}{4}+\frac{237 g_ 1^2}{80}+\frac{225 g_ 2^2}{16}+36 g_ 3^2-6 \lambda \bigg)
   \end{eqnarray}
  \begin{eqnarray}
\beta _ 2(g_ {\tau }) &=&-3 g_ {\tau }^5 + g_ {\tau } \bigg[g_ 1^2 \bigg(\frac{3 \tilde{g}_ {\text{1d}}^2}{16}+\frac{3 \tilde{g}_ {\text{1u}}^2}{16}+\frac{9 \tilde{g}_ {\text{2d}}^2}{16}+\frac{9 \tilde{g}_ {\text{2u}}^2}{16}+\frac{27 g_ 2^2}{20}\bigg)-3 \tilde{g}_ {\text{1d}} \tilde{g}_ {\text{1u}} \tilde{g}_ {\text{2d}} \tilde{g}_{\text{2u}}
+\\ \nonumber &&
+g_ 2^2 \bigg(\frac{15 \tilde{g}_ {\text{1d}}^2}{16}+\frac{15 \tilde{g}_ {\text{1u}}^2}{16}+\frac{165 \tilde{g}_ {\text{2d}}^2}{16}+\frac{165 \tilde{g}_ {\text{2u}}^2}{16}\bigg)-\frac{5}{4} \tilde{g}_ {\text{1d}}^2 \tilde{g}_ {\text{1u}}^2-\frac{9}{8} \tilde{g}_ {\text{1d}}^2 \tilde{g}_ {\text{2d}}^2-\frac{9 \tilde{g}_ {\text{1d}}^4}{16}-\frac{9}{8} \tilde{g}_ {\text{1u}}^2 \tilde{g}_ {\text{2u}}^2
+\\ \nonumber &&
-\frac{9 \tilde{g}_ {\text{1u}}^4}{16}-\frac{3}{4} \tilde{g}_ {\text{2d}}^2 \tilde{g}_ {\text{2u}}^2-\frac{45 \tilde{g}_ {\text{2d}}^4}{16}-\frac{45 \tilde{g}_ {\text{2u}}^4}{16}+\bigg(\frac{3 g_b^2}{2}+\frac{17 g_ 1^2}{8}+\frac{45 g_ 2^2}{8}+20 g_ 3^2\bigg) g_t^2
+\\ \nonumber &&
-\frac{27 g_b^4}{4}+\bigg(\frac{5 g_ 1^2}{8}+\frac{45 g_ 2^2}{8}+20 g_ 3^2\bigg) g_b^2-\frac{27 g_t^4}{4}+\frac{1503 g_ 1^4}{200}-\frac{17 g_ 2^4}{4}+\frac{3 \lambda ^2}{2}\bigg]
+\\ \nonumber &&
+g_ {\tau }^3 \bigg(-\frac{9 \tilde{g}_ {\text{1d}}^2}{8}-\frac{9 \tilde{g}_ {\text{1u}}^2}{8}-\frac{27 \tilde{g}_ {\text{2d}}^2}{8}-\frac{27 \tilde{g}_ {\text{2u}}^2}{8}-\frac{27 g_b^2}{4}-\frac{27 g_t^2}{4}+\frac{537 g_ 1^2}{80}+\frac{165 g_ 2^2}{16}-6 \lambda \bigg)
    \end{eqnarray}
The two-loop $\beta$ functions of the gaugino couplings are   
      \begin{eqnarray}
\beta _ 2(\tilde{g}_ {\text{2u}}) &=& \tilde{g}_ {\text{2u}} \bigg[g_ 1^2 \bigg(\frac{3 \tilde{g}_ {\text{1d}}^2}{16}+\frac{63 \tilde{g}_ {\text{1u}}^2}{160}+\frac{3 \tilde{g}_ {\text{2d}}^2}{20}+\frac{9 g_ 2^2}{20}\bigg)+g_ 2^2 \bigg(\frac{15 \tilde{g}_ {\text{1d}}^2}{16}+\frac{111 \tilde{g}_ {\text{1u}}^2}{32}+\frac{17 \tilde{g}_ {\text{2d}}^2}{4}\bigg)
+\\ \nonumber &&
-\frac{3}{2} \tilde{g}_ {\text{1d}}^2 \tilde{g}_ {\text{1u}}^2-\frac{13}{16} \tilde{g}_ {\text{1d}}^2 \tilde{g}_ {\text{2d}}^2-\frac{9 \tilde{g}_ {\text{1d}}^4}{16}+g_t^2 \bigg(-\frac{9 \tilde{g}_ {\text{1u}}^2}{8}+\frac{3 \tilde{g}_ {\text{2d}}^2}{4}+\frac{3 g_b^2}{2}+\frac{17 g_ 1^2}{8}+\frac{45 g_ 2^2}{8}+20 g_ 3^2\bigg)
+\\ \nonumber &&
+g_b^2 \bigg(-\frac{9 \tilde{g}_ {\text{1u}}^2}{8}+\frac{3 \tilde{g}_ {\text{2d}}^2}{4}+\frac{5 g_ 1^2}{8}+\frac{45 g_ 2^2}{8}+20 g_ 3^2\bigg)-\lambda  (\tilde{g}_ {\text{1u}}^2+\tilde{g}_ {\text{2d}}^2)
+\\ \nonumber &&
+g_ {\tau }^2 \bigg(-\frac{3 \tilde{g}_ {\text{1u}}^2}{8}+\frac{\tilde{g}_ {\text{2d}}^2}{4}+\frac{15 g_ 1^2}{8}+\frac{15 g_ 2^2}{8}\bigg)
+\\ \nonumber &&
-\frac{31}{16} \tilde{g}_ {\text{1u}}^2 \tilde{g}_ {\text{2d}}^2-\frac{5 \tilde{g}_ {\text{1u}}^4}{16}-\frac{11 \tilde{g}_ {\text{2d}}^4}{8}-\frac{27 g_b^4}{4}-\frac{27 g_t^4}{4}-\frac{9 g_ {\tau }^4}{4}
+\frac{117 g_ 1^4}{200}-\frac{409 g_ 2^4}{12}+\frac{3 \lambda ^2}{2}\bigg]
+\\ \nonumber &&
+\tilde{g}_ {\text{2u}}^3 \bigg(-\frac{15 \tilde{g}_ {\text{1d}}^2}{16}-\frac{59 \tilde{g}_ {\text{1u}}^2}{16}-\frac{27 \tilde{g}_ {\text{2d}}^2}{8}-\frac{45 g_b^2}{8}-\frac{45 g_t^2}{8}-\frac{15 g_ {\tau }^2}{8}+\frac{87 g_ 1^2}{32}+\frac{875 g_ 2^2}{32}-5 \lambda \bigg)
+\\ \nonumber &&
-4 \tilde{g}_ {\text{1d}} \tilde{g}_ {\text{1u}} \tilde{g}_ {\text{2d}} \tilde{g}_ {\text{2u}}^2-3 g_b^2 \tilde{g}_ {\text{1d}} \tilde{g}_ {\text{1u}} \tilde{g}_{\text{2d}}-\lambda  \tilde{g}_ {\text{1d}} \tilde{g}_ {\text{1u}} \tilde{g}_{\text{2d}}-3 g_t^2 \tilde{g}_ {\text{1d}} \tilde{g}_ {\text{1u}} \tilde{g}_{\text{2d}}-g_ {\tau }^2 \tilde{g}_ {\text{1d}} \tilde{g}_ {\text{1u}} \tilde{g}_{\text{2d}}
+\\ \nonumber &&
-\frac{9}{4} \tilde{g}_ {\text{1d}} \tilde{g}_ {\text{1u}} \tilde{g}_ {\text{2d}}^3-\frac{3}{2} \tilde{g}_ {\text{1d}} \tilde{g}_ {\text{1u}}^3 \tilde{g}_{\text{2d}}-\frac{5}{4} \tilde{g}_ {\text{1d}}^3 \tilde{g}_ {\text{1u}} \tilde{g}_{\text{2d}}+\frac{3}{20} g_ 1^2 \tilde{g}_ {\text{1d}} \tilde{g}_ {\text{1u}} \tilde{g}_{\text{2d}}+\frac{9}{4} g_ 2^2 \tilde{g}_ {\text{1d}} \tilde{g}_ {\text{1u}} \tilde{g}_{\text{2d}}-\frac{7 \tilde{g}_ {\text{2u}}^5}{2} 
   \end{eqnarray}
  \begin{eqnarray}
\beta _ 2(\tilde{g}_ {\text{2d}}) &=& \tilde{g}_ {\text{2d}} \bigg[g_ 1^2 \bigg(\frac{63 \tilde{g}_ {\text{1d}}^2}{160}+\frac{3 \tilde{g}_ {\text{1u}}^2}{16}+\frac{3 \tilde{g}_ {\text{2u}}^2}{20}+\frac{9 g_ 2^2}{20}\bigg)+g_ 2^2 \bigg(\frac{111 \tilde{g}_ {\text{1d}}^2}{32}+\frac{15 \tilde{g}_ {\text{1u}}^2}{16}+\frac{17 \tilde{g}_ {\text{2u}}^2}{4}\bigg)
+\\ \nonumber &&
-\frac{3}{2} \tilde{g}_ {\text{1d}}^2 \tilde{g}_ {\text{1u}}^2+g_t^2 \bigg(-\frac{9 \tilde{g}_ {\text{1d}}^2}{8}+\frac{3 \tilde{g}_ {\text{2u}}^2}{4}+\frac{3 g_b^2}{2}+\frac{17 g_ 1^2}{8}+\frac{45 g_ 2^2}{8}+20 g_ 3^2\bigg)
+\\ \nonumber &&
+g_b^2 \bigg(-\frac{9 \tilde{g}_ {\text{1d}}^2}{8}+\frac{3 \tilde{g}_ {\text{2u}}^2}{4}+\frac{5 g_ 1^2}{8}+\frac{45 g_ 2^2}{8}+20 g_ 3^2\bigg)
-\lambda  (\tilde{g}_ {\text{1d}}^2+\tilde{g}_ {\text{2u}}^2)
+\\ \nonumber &&
+g_ {\tau }^2 \bigg(-\frac{3 \tilde{g}_ {\text{1d}}^2}{8}+\frac{\tilde{g}_ {\text{2u}}^2}{4}+\frac{15 g_ 1^2}{8}+\frac{15 g_ 2^2}{8}\bigg)
-\frac{31}{16} \tilde{g}_ {\text{1d}}^2 \tilde{g}_ {\text{2u}}^2-\frac{5 \tilde{g}_ {\text{1d}}^4}{16}
+\\ \nonumber &&
-\frac{13}{16} \tilde{g}_ {\text{1u}}^2 \tilde{g}_ {\text{2u}}^2-\frac{9 \tilde{g}_ {\text{1u}}^4}{16}-\frac{11 \tilde{g}_ {\text{2u}}^4}{8}-\frac{27 g_b^4}{4}-\frac{27 g_t^4}{4}-\frac{9 g_ {\tau }^4}{4}+\frac{117 g_ 1^4}{200}-\frac{409 g_ 2^4}{12}+\frac{3 \lambda ^2}{2}\bigg]
+\\ \nonumber &&
+\tilde{g}_ {\text{2d}}^3 \bigg(-\frac{59 \tilde{g}_ {\text{1d}}^2}{16}-\frac{15 \tilde{g}_ {\text{1u}}^2}{16}-\frac{27 \tilde{g}_ {\text{2u}}^2}{8}-\frac{45 g_b^2}{8}-\frac{45 g_t^2}{8}-\frac{15 g_ {\tau }^2}{8}+\frac{87 g_ 1^2}{32}+\frac{875 g_ 2^2}{32}-5 \lambda \bigg)
+\\ \nonumber &&
-4 \tilde{g}_ {\text{1d}} \tilde{g}_ {\text{1u}} \tilde{g}_ {\text{2d}}^2 \tilde{g}_{\text{2u}}-3 g_b^2 \tilde{g}_ {\text{1d}} \tilde{g}_ {\text{1u}} \tilde{g}_{\text{2u}}-\lambda  \tilde{g}_ {\text{1d}} \tilde{g}_ {\text{1u}} \tilde{g}_{\text{2u}}-3 g_t^2 \tilde{g}_ {\text{1d}} \tilde{g}_ {\text{1u}} \tilde{g}_{\text{2u}}-g_ {\tau }^2 \tilde{g}_ {\text{1d}} \tilde{g}_ {\text{1u}} \tilde{g}_{\text{2u}}
+\\ \nonumber &&
-\frac{9}{4} \tilde{g}_ {\text{1d}} \tilde{g}_ {\text{1u}} \tilde{g}_ {\text{2u}}^3-\frac{5}{4} \tilde{g}_ {\text{1d}} \tilde{g}_ {\text{1u}}^3 \tilde{g}_{\text{2u}}-\frac{3}{2} \tilde{g}_ {\text{1d}}^3 \tilde{g}_ {\text{1u}} \tilde{g}_{\text{2u}}+\frac{3}{20} g_ 1^2 \tilde{g}_ {\text{1d}} \tilde{g}_ {\text{1u}} \tilde{g}_{\text{2u}}+\frac{9}{4} g_ 2^2 \tilde{g}_ {\text{1d}} \tilde{g}_ {\text{1u}} \tilde{g}_{\text{2u}}-\frac{7 \tilde{g}_ {\text{2d}}^5}{2} 
    \end{eqnarray}
       \begin{eqnarray}
\beta _ 2(\tilde{g}_ {\text{1u}}) &=& \tilde{g}_ {\text{1u}} 
\bigg[g_ 1^2 \bigg(\frac{3 \tilde{g}_ {\text{1d}}^2}{40}+\frac{9 \tilde{g}_ {\text{2d}}^2}{16}+\frac{189 \tilde{g}_ {\text{2u}}^2}{160}-\frac{27 g_ 2^2}{20}\bigg)+
g_ 2^2 \bigg(\frac{39 \tilde{g}_ {\text{1d}}^2}{8}+\frac{165 \tilde{g}_ {\text{2d}}^2}{16}+\frac{549 \tilde{g}_ {\text{2u}}^2}{32}\bigg)
+\\ \nonumber &&
-\frac{75}{16} \tilde{g}_ {\text{1d}}^2 \tilde{g}_ {\text{2d}}^2+g_t^2 \bigg(-\frac{21 \tilde{g}_ {\text{1d}}^2}{4}-\frac{27 \tilde{g}_ {\text{2u}}^2}{8}+\frac{3 g_b^2}{2}+\frac{17 g_ 1^2}{8}+\frac{45 g_ 2^2}{8}+20 g_ 3^2\bigg)
+\\ \nonumber &&
+g_b^2 \bigg(-\frac{21 \tilde{g}_ {\text{1d}}^2}{4}-\frac{27 \tilde{g}_ {\text{2u}}^2}{8}+\frac{5 g_ 1^2}{8}+\frac{45 g_ 2^2}{8}+20 g_ 3^2\bigg)-
3\lambda  ( \tilde{g}_ {\text{1d}}^2+ \tilde{g}_ {\text{2u}}^2)
+\\ \nonumber &&
+g_ {\tau }^2 \bigg(-\frac{7 \tilde{g}_ {\text{1d}}^2}{4}-\frac{9 \tilde{g}_ {\text{2u}}^2}{8}+\frac{15 g_ 1^2}{8}+\frac{15 g_ 2^2}{8}\bigg)
-\frac{75}{16} \tilde{g}_ {\text{1d}}^2 \tilde{g}_ {\text{2u}}^2-\frac{9 \tilde{g}_ {\text{1d}}^4}{4}
+\\ \nonumber &&
-\frac{21}{8} \tilde{g}_ {\text{2d}}^2 \tilde{g}_ {\text{2u}}^2-\frac{45 \tilde{g}_ {\text{2d}}^4}{16}-\frac{99 \tilde{g}_ {\text{2u}}^4}{16}-\frac{27 g_b^4}{4}-\frac{27 g_t^4}{4}-\frac{9 g_ {\tau }^4}{4}+\frac{117 g_ 1^4}{200}-\frac{17 g_ 2^4}{4}+\frac{3 \lambda ^2}{2}\bigg]
+\\ \nonumber &&
+\tilde{g}_ {\text{1u}}^3 \bigg(-\frac{15 \tilde{g}_ {\text{1d}}^2}{4}-\frac{27 \tilde{g}_ {\text{2d}}^2}{16}-\frac{9 \tilde{g}_ {\text{2u}}^2}{16}-\frac{27 g_b^2}{8}-\frac{27 g_t^2}{8}-\frac{9 g_ {\tau }^2}{8}+\frac{309 g_ 1^2}{160}+\frac{165 g_ 2^2}{32}-3 \lambda \bigg)
+\\ \nonumber &&
-6 \tilde{g}_ {\text{1d}} \tilde{g}_ {\text{1u}}^2 \tilde{g}_ {\text{2d}} \tilde{g}_{\text{2u}}-9 g_b^2 \tilde{g}_ {\text{1d}} \tilde{g}_ {\text{2d}} \tilde{g}_{\text{2u}}-3 \lambda  \tilde{g}_ {\text{1d}} \tilde{g}_ {\text{2d}} \tilde{g}_{\text{2u}}-9 g_t^2 \tilde{g}_ {\text{1d}} \tilde{g}_ {\text{2d}} \tilde{g}_{\text{2u}}-3 g_ {\tau }^2 \tilde{g}_ {\text{1d}} \tilde{g}_ {\text{2d}} \tilde{g}_{\text{2u}}
+\\ \nonumber &&
-\frac{9}{2} \tilde{g}_ {\text{1d}} \tilde{g}_ {\text{2d}} \tilde{g}_ {\text{2u}}^3-\frac{33}{4} \tilde{g}_ {\text{1d}} \tilde{g}_ {\text{2d}}^3 \tilde{g}_{\text{2u}}-\frac{9}{4} \tilde{g}_ {\text{1d}}^3 \tilde{g}_ {\text{2d}} \tilde{g}_{\text{2u}}+\frac{9}{20} g_ 1^2 \tilde{g}_ {\text{1d}} \tilde{g}_ {\text{2d}} \tilde{g}_{\text{2u}}+\frac{51}{4} g_ 2^2 \tilde{g}_ {\text{1d}} \tilde{g}_ {\text{2d}} \tilde{g}_{\text{2u}}-\frac{3 \tilde{g}_ {\text{1u}}^5}{4} 
   \end{eqnarray}
  \begin{eqnarray}
\beta _ 2(\tilde{g}_ {\text{1d}}) &=& \tilde{g}_ {\text{1d}} \bigg[g_ 1^2 \bigg(\frac{3 \tilde{g}_ {\text{1u}}^2}{40}+\frac{189 \tilde{g}_ {\text{2d}}^2}{160}+\frac{9 \tilde{g}_ {\text{2u}}^2}{16}-\frac{27 g_ 2^2}{20}\bigg)+g_ 2^2 \bigg(\frac{39 \tilde{g}_ {\text{1u}}^2}{8}+\frac{549 \tilde{g}_ {\text{2d}}^2}{32}+\frac{165 \tilde{g}_ {\text{2u}}^2}{16}\bigg)
+\\ \nonumber &&
+g_t^2 \bigg(-\frac{21 \tilde{g}_ {\text{1u}}^2}{4}-\frac{27 \tilde{g}_ {\text{2d}}^2}{8}+\frac{3 g_b^2}{2}+\frac{17 g_ 1^2}{8}+\frac{45 g_ 2^2}{8}+20 g_ 3^2\bigg)+\\ \nonumber &&
+g_b^2 \bigg(-\frac{21 \tilde{g}_ {\text{1u}}^2}{4}-\frac{27 \tilde{g}_ {\text{2d}}^2}{8}+\frac{5 g_ 1^2}{8}+\frac{45 g_ 2^2}{8}+20 g_ 3^2\bigg)-3\lambda  ( \tilde{g}_ {\text{1u}}^2+ \tilde{g}_ {\text{2d}}^2)
+\\ \nonumber &&
+g_ {\tau }^2 \bigg(-\frac{7 \tilde{g}_ {\text{1u}}^2}{4}-\frac{9 \tilde{g}_ {\text{2d}}^2}{8}+\frac{15 g_ 1^2}{8}+\frac{15 g_ 2^2}{8}\bigg)-\frac{75}{16} \tilde{g}_ {\text{1u}}^2 \tilde{g}_ {\text{2d}}^2-\frac{75}{16} \tilde{g}_ {\text{1u}}^2 \tilde{g}_ {\text{2u}}^2-\frac{9 \tilde{g}_ {\text{1u}}^4}{4}
+\\ \nonumber &&
-\frac{21}{8} \tilde{g}_ {\text{2d}}^2 \tilde{g}_ {\text{2u}}^2-\frac{99 \tilde{g}_ {\text{2d}}^4}{16}-\frac{45 \tilde{g}_ {\text{2u}}^4}{16}-\frac{27 g_b^4}{4}-\frac{27 g_t^4}{4}-\frac{9 g_ {\tau }^4}{4}+\frac{117 g_ 1^4}{200}-\frac{17 g_ 2^4}{4}+\frac{3 \lambda ^2}{2}\bigg]
+\\ \nonumber &&
+\tilde{g}_ {\text{1d}}^3 \bigg(-\frac{15 \tilde{g}_ {\text{1u}}^2}{4}-\frac{9 \tilde{g}_ {\text{2d}}^2}{16}-\frac{27 \tilde{g}_ {\text{2u}}^2}{16}-\frac{27 g_b^2}{8}-\frac{27 g_t^2}{8}-\frac{9 g_ {\tau }^2}{8}+\frac{309 g_ 1^2}{160}+\frac{165 g_ 2^2}{32}-3 \lambda \bigg)
+\\ \nonumber &&
-6 \tilde{g}_ {\text{1d}}^2 \tilde{g}_ {\text{1u}} \tilde{g}_ {\text{2d}} \tilde{g}_{\text{2u}}-\frac{3 \tilde{g}_ {\text{1d}}^5}{4}-9 g_b^2 \tilde{g}_ {\text{1u}} \tilde{g}_ {\text{2d}} \tilde{g}_{\text{2u}}-3 \lambda  \tilde{g}_ {\text{1u}} \tilde{g}_ {\text{2d}} \tilde{g}_{\text{2u}}-9 g_t^2 \tilde{g}_ {\text{1u}} \tilde{g}_ {\text{2d}} \tilde{g}_{\text{2u}}-3 g_ {\tau }^2 \tilde{g}_ {\text{1u}} \tilde{g}_ {\text{2d}} \tilde{g}_{\text{2u}}
+\\ \nonumber &&
-\frac{33}{4} \tilde{g}_ {\text{1u}} \tilde{g}_ {\text{2d}} \tilde{g}_ {\text{2u}}^3-\frac{9}{2} \tilde{g}_ {\text{1u}} \tilde{g}_ {\text{2d}}^3 \tilde{g}_{\text{2u}}-\frac{9}{4} \tilde{g}_ {\text{1u}}^3 \tilde{g}_ {\text{2d}} \tilde{g}_{\text{2u}}+\frac{9}{20} g_ 1^2 \tilde{g}_ {\text{1u}} \tilde{g}_ {\text{2d}} \tilde{g}_{\text{2u}}+\frac{51}{4} g_ 2^2 \tilde{g}_ {\text{1u}} \tilde{g}_ {\text{2d}} \tilde{g}_{\text{2u}} 
    \end{eqnarray}   
     The two-loop $\beta$ functions of quartic Higgs coupling is 
       \begin{eqnarray}\label{lambda2loop}
\beta _ 2(\lambda ) &=& -78 \lambda ^3-\frac{3699 g_ 1^6}{1000}+\bigg(\frac{9 g_b^2}{10}-\frac{171 g_t^2}{50}-\frac{9 g_ {\tau }^2}{2}-\frac{9}{100} (\tilde{g}_ {\text{1d}}^2+\tilde{g}_ {\text{1u}}^2+3 \tilde{g}_ {\text{2d}}^2+3 \tilde{g}_ {\text{2u}}^2)\bigg) g_ 1^4
+\\ \nonumber &&
+\bigg(\frac{8 g_b^4}{5}-\frac{16 g_t^4}{5}-\frac{24 g_ {\tau }^4}{5}\bigg) g_ 1^2+\frac{209 g_ 2^6}{8}+20 (3g_b^6+3 g_t^6+ g_ {\tau }^6)-64 g_ 3^2 g_b^4
-
(64 g_ 3^2+12 g_b^2) g_t^4
+\\ \nonumber &&
-12 g_b^4 g_t^2+\lambda ^2 \bigg(\frac{54 g_ 1^2}{5}+54 g_ 2^2-72 g_b^2-72 g_t^2-24 g_ {\tau }^2-12 \bigg(\tilde{g}_ {\text{1d}}^2+\tilde{g}_ {\text{1u}}^2+3 \tilde{g}_ {\text{2d}}^2+3 \tilde{g}_ {\text{2u}}^2\bigg)\bigg)
+\\ \nonumber &&
+g_ 2^4 \bigg(-\frac{77 g_ 1^2}{8}-\frac{9 g_b^2}{2}-\frac{9 g_t^2}{2}-\frac{3 g_ {\tau }^2}{2}-\frac{3}{4} (\tilde{g}_ {\text{1d}}^2+\tilde{g}_ {\text{1u}}^2+51 \tilde{g}_ {\text{2d}}^2+51 \tilde{g}_ {\text{2u}}^2)\bigg)
+\\ \nonumber &&
+g_ 2^2 \bigg[-\frac{1773 g_ 1^4}{200}+\bigg(\frac{27 g_b^2}{5}+\frac{63 g_t^2}{5}+\frac{33 g_ {\tau }^2}{5}-\frac{3}{10} (\tilde{g}_ {\text{1d}}^2+\tilde{g}_ {\text{1u}}^2-21 \tilde{g}_ {\text{2d}}^2-21 \tilde{g}_ {\text{2u}}^2)\bigg) g_ 1^2
+\\ \nonumber &&
-4 \bigg(5 \tilde{g}_ {\text{2d}}^4+\tilde{g}_ {\text{1d}}^2 \tilde{g}_ {\text{2d}}^2+2 \tilde{g}_ {\text{2u}}^2 \tilde{g}_ {\text{2d}}^2+2 \tilde{g}_ {\text{1d}} \tilde{g}_ {\text{1u}} \tilde{g}_ {\text{2u}} \tilde{g}_{\text{2d}}+5 \tilde{g}_ {\text{2u}}^4+\tilde{g}_ {\text{1u}}^2 \tilde{g}_ {\text{2u}}^2\bigg)\bigg]
+\\ \nonumber &&
+\lambda  \bigg[\frac{2007 g_ 1^4}{200}+\bigg(\frac{5 g_b^2}{2}+\frac{17 g_t^2}{2}+\frac{15 g_ {\tau }^2}{2}+\frac{3}{4} (\tilde{g}_ {\text{1d}}^2+\tilde{g}_ {\text{1u}}^2+3 \tilde{g}_ {\text{2d}}^2+3 \tilde{g}_ {\text{2u}}^2)\bigg) g_ 1^2+\frac{47 g_ 2^4}{8}
+\\ \nonumber &&
-3 g_b^4-3 g_t^4-g_ {\tau }^4+80 g_ 3^2 g_b^2+(80 g_ 3^2-42 g_b^2) g_t^2
+\\ \nonumber &&
+\frac{1}{4} \bigg(-\tilde{g}_ {\text{1d}}^4-2 (\tilde{g}_ {\text{2d}}^2-6 \tilde{g}_ {\text{1u}}^2) \tilde{g}_ {\text{1d}}^2+80 \tilde{g}_ {\text{1u}} \tilde{g}_ {\text{2d}} \tilde{g}_ {\text{2u}} \tilde{g}_{\text{1d}}-\tilde{g}_ {\text{1u}}^4-5 \tilde{g}_ {\text{2d}}^4-5 \tilde{g}_ {\text{2u}}^4-2 \tilde{g}_ {\text{1u}}^2 \tilde{g}_ {\text{2u}}^2-44 \tilde{g}_ {\text{2d}}^2 \tilde{g}_ {\text{2u}}^2\bigg)
+\\ \nonumber &&
+g_ 2^2 \bigg(\frac{117 g_ 1^2}{20}+\frac{45 g_b^2}{2}+\frac{45 g_t^2}{2}+\frac{15 g_ {\tau }^2}{2}+\frac{15}{4} (\tilde{g}_ {\text{1d}}^2+\tilde{g}_ {\text{1u}}^2+11 \tilde{g}_ {\text{2d}}^2+11 \tilde{g}_ {\text{2u}}^2)\bigg)\bigg]
+\\ \nonumber &&
+\frac{1}{2} \bigg[
 47 (\tilde{g}_{\text{2d}}^6+
    \tilde{g}_{\text{2u}}^6)+
    5 (\tilde{g}_{\text{1d}}^6+
    \tilde{g}_{\text{1u}}^6)+
\tilde{g}_{\text{1d}} \tilde{g}_{\text{1u}} \tilde{g}_{\text{2d}} \tilde{g}_{\text{2u}}
(42\tilde{g}_{\text{1d}}^2+42\tilde{g}_{\text{1u}}^2+
38  \tilde{g}_{\text{2u}}^2+
   38 \tilde{g}_{\text{2d}}^2 )
   +\\ \nonumber &&
   19\tilde{g}_{\text{1d}}^2 \tilde{g}_{\text{1u}}^2 (\tilde{g}_{\text{2d}}^2+
    \tilde{g}_{\text{2u}}^2)+
21  \tilde{g}_{\text{2d}}^2  \tilde{g}_{\text{2u}}^2(  \tilde{g}_{\text{1d}}^2+ \tilde{g}_{\text{1u}}^2)+
     17 (\tilde{g}_{\text{1d}}^4 \tilde{g}_{\text{1u}}^2+
   \tilde{g}_{\text{1d}}^2 \tilde{g}_{\text{1u}}^4+
    \tilde{g}_{\text{1u}}^4  \tilde{g}_{\text{2u}}^2+
   \tilde{g}_{\text{1d}}^4 \tilde{g}_{\text{2d}}^2)
      +\\ \nonumber &&
      +
    11 (\tilde{g}_{\text{1d}}^2   \tilde{g}_{\text{2d}}^4+
    \tilde{g}_{\text{1u}}^2 \tilde{g}_{\text{2u}}^4)+
    7 \tilde{g}_{\text{2d}}^2 \tilde{g}_{\text{2u}}^2( \tilde{g}_{\text{2u}}^2+ \tilde{g}_{\text{2d}}^2)
   \bigg]
      \end{eqnarray}      
      The two-loop RGE for $\lambda$ has been previously computed in ref.~\cite{Binger}.\footnote{Some terms
      containing $g_2$ were incorrect in the previous version of this paper, see~\cite{Tamarit}.}

\section{Thresholds at the weak scale}\label{soglie}
The functions that enter the SM weak threshold corrections to the Higgs mass are:
\begin{eqnsystem}{sys:F}
F_1 &=& 12\ln\frac{Q}{m_h}+\frac{3}{2}\ln\xi - \frac{1}{2}
Z(\frac{1}{\xi}) -Z(\frac{c_W^2}{\xi})-\ln c_W^2+\frac{9}{2} (\frac{25}{9}-\frac{\pi}{\sqrt{3}})\\
F_0 &=& -12\ln\frac{Q}{M_Z}(1+2c_W^2-2 \frac{m_t^2}{M_Z^2}) + \frac{3c_W^2\xi}{\xi-c_W^2}\ln\frac{\xi}{c_W^2}+2Z(\frac{1}{\xi}) +4c_W^2Z(\frac{c_W^2}{\xi})
+3\frac{c_W^2}{s_W^2}\ln c_W^2+
\nonumber   \\ &&
+12 c_W^2 \ln c_W^2-\frac{15}{2}(1+2c_W^2)-3\frac{m_T^2}{M_Z^2}[2 Z(\frac{m_t^2}{M_Z^2 \xi})-5+4\ln\frac{m_t^2}{M_Z^2}] \\
F_3 &=& 12\ln\frac{Q}{M_Z}(1+2c_W^4-4 \frac{m_t^4}{M_Z^4}) -6 Z(\frac{1}{\xi})+
\nonumber   \\ &&
 -12c_W^4  Z(\frac{c_W^2}{\xi})-12 c_W^4\ln c_W^2+8
(1+2c_W^4) + 24 \frac{m_t^4}{M_Z^4}[Z(\frac{m_t^2}{M_Z^2 \xi})-2+\ln\frac{m_t^2}{M_Z^2} ] 
\end{eqnsystem}
where $c_W =\cos\theta_W$, $s_W = \sin\theta_W$, $\xi = m_h^2/M_Z^2$ and 
\beq Z(z) =\left\{\begin{array}{ll} 
2\zeta \arctan(1/\zeta) & \hbox{for $z>1/4$}\\
\zeta \ln[(1+\zeta)/(1-\zeta)] & \hbox{for $z<1/4$}
\end{array}\right.\qquad\hbox{where}\quad \zeta=\sqrt{|1-4z|} .\eeq

The functions that enter the Split Supersymmetry weak thresholds are:
\begin{eqnsystem}{sys:f}
f(r) &=& \frac{3r(r^2+1)}{(r^2-1)^2}-\frac{12r^3\ln r}{(r^2-1)^3}\\
g(r)&=&-\frac{3(r^4-6r^2+1)}{2(r^2-1)^2} + \frac{6r^4(r^2-3)\ln r}{(r^2-1)^3}\\
f_1(r) &=& \frac{6 \left(r^2+3\right) r^2}{7 \left(r^2-1\right)^2}+\frac{12 \left(r^2-5\right) r^4 \ln r}{7 \left(r^2-1\right)^3}\\
f_2(r) &=& \frac{2 \left(r^2+11\right) r^2}{9 \left(r^2-1\right)^2}+\frac{4 \left(5 r^2-17\right) r^4 \ln r}{9 \left(r^2-1\right)^3}\\
f_3(r) &=&\frac{2 \left(r^4+9 r^2+2\right)}{3 \left(r^2-1\right)^2}+\frac{4 \left(r^4-7 r^2-6\right) r^2 \ln r}{3 \left(r^2-1\right)^3}\\
 f_4(r)&=&\frac{2 \left(5 r^4+25 r^2+6\right)}{7 \left(r^2-1\right)^2}+\frac{4 \left(r^4-19 r^2-18\right) r^2 \ln r}{7 \left(r^2-1\right)^3}\\
\frac{4}{3}  f_5(r_1,r_2) &=&\frac{1+(r_1+r_2)^2- r_1^2 r_2^2}{\left(r_1^2-1\right) \left(r_2^2-1\right)}+\frac{2r_1^3 \left(r_1^2+1\right)  \ln 
   r_1}{\left(r_1^2-1\right){}^2 \left(r_1-r_2\right)}-\frac{2 r_2^3 \left(r_2^2+1\right) \ln 
   r_2}{\left(r_1-r_2\right) \left(r_2^2-1\right){}^2}\\
    \frac{7}{6}  f_6 (r_1,r_2)  &=& \frac{r_1^2+r_2^2+r_1 r_2-r_1^2 r_2^2}{\left(r_1^2-1\right)  \left(r_2^2-1\right)}+\frac{2 r_1^5 \ln 
r_1 }{\left(r_1^2-1\right){}^2  \left(r_1-r_2\right)}-\frac{2 r_2^5 \ln 
r_2}{\left(r_1-r_2\right) \left(r_2^2-1\right){}^2 }\\
\frac{1}{6}f_7(r_1,r_2) &=& \frac{1+r_1 r_2}{\left(r_1^2-1\right) \left(r_2^2-1\right)}+\frac{2 r_1^3 \ln  r_1}{\left(r_1^2-1\right){}^2
   \left(r_1-r_2\right)}-\frac{2 r_2^3 \ln  r_2}{\left(r_1-r_2\right) \left(r_2^2-1\right){}^2}\\
  \frac{2}{3} f_8(r_1,r_2)  &=& \frac{r_1+r_2}{\left(r_1^2-1\right) \left(r_2^2-1\right)}+\frac{2 r_1^4 \ln  r_1}{\left(r_1^2-1\right){}^2
   \left(r_1-r_2\right)}-\frac{2 r_2^4 \ln  r_2}{\left(r_1-r_2\right) \left(r_2^2-1\right){}^2} .
\end{eqnsystem}
All these functions are equal to 1 when they arguments approach unity.

\bigskip

\def\bf{\rm}

\small
\begin{multicols}{2}

\end{multicols}

\begin{thebibliography}{nn}

  \bibitem{Hisano}
\art[hep-ph/0610249]{J. Hisano, S. Matsumoto, M. Nagai, O. Saito, M. Senami}{Phys. Lett.}{B646}{34}{2007}.
\art[0706.4071]{M. Cirelli, A. Strumia, M. Tamburini}{Nucl. Phys.}{B787}{2007}{152}.

\bibitem{ftlhc}
\art[1101.2195]{A. Strumia}{JHEP}{1104}{2011}{073}.

\bibitem{split}
  N.~Arkani-Hamed and S.~Dimopoulos,
  JHEP {\bf 0506} (2005) 073
  [arXiv:hep-th/0405159].

  
  
  \bibitem{GR}     G.~F.~Giudice and A.~Romanino,
  Nucl.\ Phys.\  B {\bf 699} (2004) 65
  [arXiv:hep-ph/0406088].
  

  
  \bibitem{GR4}
      N.~Arkani-Hamed, S.~Dimopoulos, G.~F.~Giudice and A.~Romanino,
  Nucl.\ Phys.\  B {\bf 709} (2005) 3
  [arXiv:hep-ph/0409232].

\bibitem{Binger}
\art[hep-ph/0408240]{M. Binger}{Phys. Rev.}{D73}{095001}{2006}.



\bibitem{new}
  D.~S.~M.~Alves, E.~Izaguirre and J.~G.~Wacker,
  arXiv:1108.3390;
  M.~E.~Cabrera, J.~A.~Casas and A.~Delgado,
  arXiv:1108.3867.

\bibitem{nomura}
  L.~J.~Hall and Y.~Nomura,
  JHEP {\bf 1003} (2010) 076
  [arXiv:0910.2235].
  
    \bibitem{Slavich14}
E.~Bagnaschi, G.~F.~Giudice, P.~Slavich and A.~Strumia,
  arXiv:\hhref{1407.4081}.

\bibitem{Mah}
\hepart[hep-ph/0408096]{R.  Mahbubani}.

\bibitem{LG}
\hepart[1112.2635]{P. Lodone, P.P. Giardino}.

\bibitem{Sirlin} A. Sirlin and R. Zucchini, Nucl. Phys. B 266 (1986) 389.


\bibitem{deltat}
K.G. Chetyrkin, M. Steinhauser, Nucl. Phys. B573 (2000) 617 [arXiv:hep-ph/9911434].

\bibitem{Giardino}
D.~Buttazzo et al.,
  JHEP {1312} (2013) 089
  [arXiv:\hhref{1307.3536}].

\bibitem{topmass}
\hepart[1107.5255]{Tevatron Electroweak Working Group}.

\bibitem{alpha3} S. Bethke, arXiv:0908.1135.


\bibitem{Slavich} 
\art[0705.1496]{N. Bernal, A. Djouadi, P. Slavich}{JHEP}{0707}{016}{2007}.


\bibitem{vacdecay}
S.~Coleman, Phys. Rev. D. {15 (1977) 2929};
\art[hep-ph/0104016]{G.~Isidori, G.~Ridolfi and A.~Strumia}{Nucl.\ Phys.}{B609}{287}{2001};
J.~R.~Espinosa, G.~F.~Giudice and A.~Riotto,
  JCAP {\bf 0805} (2008) 002
  [arXiv:0710.2484];
\art[0712.0242]{G. Isidori et al.}{Phys. Rev.}{D77}{25034}{2008}.

\bibitem{mhexp}
\href{http://indico.cern.ch/conferenceDisplay.py?confId=164890}{Update on the Standard Model Higgs searches in ATLAS and CMS},
talks by F. Gianotti  and G. Tonelli, 13/12/2011, CERN.

\bibitem{slavich}
  P.~Gambino, G.~F.~Giudice and P.~Slavich,
  Nucl.\ Phys.\ B {\bf 726} (2005) 35
  [hep-ph/0506214].
  
  \bibitem{Tamarit}
  C. Tamarit, arXiv:\hhref{1204.2292}.
   K.~Benakli, L.~DarmŽ, M.~D.~Goodsell and P.~Slavich,
  JHEP {1405} (2014) 113
  [arXiv:\hhref{1312.5220}].


\bibitem{MV}   M.E. Machacek and M.T. Vaughn, Nucl. Phys. B222 (1983) 83;
M.E. Machacek and M.T. Vaughn, Nucl. Phys. B236 (1984)  221;
M.E. Machacek and M.T. Vaughn, Nucl. Phys. B249 (1985) 70.

\bibitem{luo}
\art[hep-ph/0207271]{M. Luo, Y. Xiao}{Phys. Rev. Lett.}{90}{011601}{2003};
\art[hep-ph/0211440]{M. Luo, H. Wang, Y. Xiao}{Phys. Rev.}{D67}{065019}{2003}.



\end{thebibliography}
\end{document}